\documentclass[aps,pre,twocolumn,showpacs,normalem]{revtex4}
\pdfoutput=1

\usepackage{amsmath,amssymb}
\usepackage{bm}
\usepackage{graphicx}

\usepackage{color}

\usepackage{ulem}
\newcommand{\revision}[1]{\textcolor{black}{#1}}

\begin{document}

\title{
Hydrodynamic description of (visco)elastic composite materials \revision{and} relative strains as a new macroscopic variable
}

\author{Andreas M.\ Menzel}
\email{menzel@thphy.uni-duesseldorf.de}
\affiliation{Institut f\"ur Theoretische Physik II: Weiche Materie, Heinrich-Heine-Universit\"at D\"usseldorf, D-40225 D\"usseldorf, Germany
}

\date{\today}

\begin{abstract}
One possibility to adjust material properties to a specific need is to embed units of one substance into a matrix of another substance. Even materials that are readily tunable during operation can be generated in this way. In (visco)elastic substances, both the matrix material as well as the inclusions and/or their immediate environment can be dynamically deformed. If the typical dynamic response time of the inclusions and their surroundings approach the macroscopic response time, their deformation processes need to be included into a dynamic macroscopic characterization. Along these lines, we present a hydrodynamic description of (visco)elastic composite materials. For this purpose, additional strain variables reflect the state of the inclusions and their immediate environment. These additional strain variables in general are not set by a coarse-grained macroscopic displacement field. Apart from that, during our derivation, we also include the macroscopic variables of relative translations and relative rotations that were previously introduced in different contexts. \revision{As a central point, our approach reveals and classifies the importance of a new macroscopic variable: \textit{relative strains}.} We analyze two simplified minimal example geometries as an illustration.  
\end{abstract}














\maketitle


\section{Introduction} \label{sec:introduction}

Concrete is regarded as the most-used fabricated material on earth \cite{kurtis2015innovations}. It is a composite material that consists of particulate inclusion material stuck together by a cement matrix \cite{landis2009explicit,kurtis2015innovations}. A carefully selected choice of the inclusions serves to adjust the material properties to the current need. The amount of possible combinations appears vast. For instance, concrete reinforcement using vegetable fibers is discussed, motivated for instance by environmental reasons \cite{pacheco2011cementitious}. The same principles of generating composite materials of new adjusted properties can of course be transferred to different classes of substances. For example, flexible polymer materials can be reinforced using carbon nanotubes \cite{coleman2006mechanical}. 
In a further step, inclusions may be added that can be selectively addressed by external fields to reversibly adjust the material properties during operation. Examples are magnetic colloidal particles in a gel matrix, which allows to control the elastic properties by external magnetic fields \cite{filipcsei2007magnetic,menzel2015tuned,odenbach2016microstructure}.

Our focus in this manuscript is mainly on elastic and viscoelastic (quasi-)static and dynamic properties of such composite materials, ignoring at this time electric and magnetic field effects. Yet, couplings to other variables are taken into account. 
For this purpose, we use and extend a macroscopic hydrodynamic 
approach that is 
based on symmetry arguments \cite{martin1972unified,pleiner1996pattern}. Strictly speaking, this hydrodynamic framework applies to macroscopic variables that do not relax in the spatially homogeneous limit. This is true for quantities characterized by local conservation laws, e.g.\ mass, momentum, and energy \cite{kadanoff1963hydrodynamic,forster1975hydrodynamic,landau1987fluid}, as well as for variables describing spontaneously broken continuous symmetries, such as the nematic director in nematic liquid crystals \cite{forster1975hydrodynamic,degennes1993physics}. In addition to that, relaxation processes described by other variables and degrees of freedom may reach time scales comparable to the hydrodynamic one. Then, they may influence or interfere with the macroscopic dynamics. In such situations, they should be included on a symmetry basis into the macroscopic description as so-called slowly relaxing variables \cite{pleiner1996pattern}. 

In this study, we consider materials composed of more or less elastic or viscoelastic inclusions embedded in an elastic or viscoelastic matrix. At first glance, it seems a little contradictory that the dynamics of small embedded inclusions should reach the macroscopic time scales of the overall material. Yet, for instance in polymeric materials, such situations are conceivable. For example, a matrix of permanently crosslinked polymer can respond in a relatively quick and elastic way \cite{treloar1975physics,strobl1997physics}. In contrast to that, the relaxation of non-permanently crosslinked but strongly entangled polymer inclusions may reach macroscopic time scales \cite{degennes1979scaling,strobl1997physics,doi2007theory}. 
Likewise, our theory should be applicable to describe dynamic aspects of different kinds of interpenetrating polymer networks \cite{sperling1977interpenetrating,suthar1996review,sperling1996current,myung2008progress}. 

Below, we will distinguish between three instances of (visco)elastic strain deformation that we here refer to in the following way: the matrix, the inclusions, and a coupling zone between the two. 
An inspiration to include an additional coupling zone separately came from an experimental observation on magnetic elastomers, i.e.\ rigid magnetic colloidal particles embedded in a crosslinked polymer matrix \cite{filipcsei2007magnetic,menzel2015tuned,odenbach2016microstructure}. From x-ray microtomographic investigations, it was concluded that the polymer in the close vicinity of the particles got significantly less crosslinked \cite{gundermann2014investigation}. Also the opposite case of a stiffer immediate particle environment can be observed \cite{huang2016buckling}. Such deviations from the bulk properties can be described by including coupling zones. Coming back to our introductory example, coupling zones are likewise encountered in concrete. Here, the cement matrix in the direct vicinity of the inclusions is observed to show a porosity different from its bulk value \cite{landis2009explicit}. 

Other systems that explicitly show the structure referred to above are block copolymer melts and solutions in their microphase-separated states \cite{hamley1998physics,park2003enabling,ohta1986equilibrium, nakazawa1993microphase,liedel2012beyond}. Linear triblock copolymers are made of linear polymer chains that exhibit three different blocks. These blocks can feature different chemical properties \cite{park2003enabling,mogi1992preparation,nakazawa1993microphase}. Under microphase separation, blocks of identical chemical nature tend to gather and to avoid the other blocks. However, a macroscopic phase separation is not possible. On each chain, the different blocks are linked to each other at their ends. From this frustration, regularly arranged domains of the dimension of the block size result. In fact, in a certain microphase-separated state, a continuous matrix can be formed by the blocks on one end of the linear chains. The blocks on the other end of the polymer chains can be regularly arranged within this matrix in the form of spherical inclusions. Finally, the central blocks on the polymer chains can compose shells around these inclusions, coupling the inclusions to the matrix \cite{park2003enabling}. Due to different chemical properties of the different blocks, the three zones can feature different elastic and relaxation behaviors. These and further microphase-separated states \cite{hamley1998physics,park2003enabling,ohta1986equilibrium, mogi1992preparation,nakazawa1993microphase,liedel2012beyond,menzel2015tuned}, e.g.\ regular lamellar phases, of these and other materials 
may likewise be covered by variants of our description. 

In part, these are very specific systems. Yet, our approach is based on symmetry arguments and can be transferred to any other material featuring similar prerequisites. If any of the three levels mentioned above does not play a role for the macroscopic system dynamics, e.g.\ strain deformations of rigid inclusions, its influence can simply be discarded. 

Thus, as a central point of our description, the different components do not have to deform in the same way, but may deform differently with respect to each other. This leads to \textit{relative strains} between the different components. As we will demonstrate towards the end, the whole theory can be set up by initially using relative strains as macroscopic variables instead of different strains for different components. In this way, relative strains are introduced as a new macroscopic variable. They complement two previous hydrodynamic concepts introduced in different contexts: on the one hand \textit{relative translations}, applied in the context of incommensurate smectic phases and chiral smectic polymers \cite{brand1983hydrodynamics,pleiner1992macroscopic}; on the other hand \textit{relative rotations}, introduced to characterize nematic and cholesteric liquid crystalline elastomers \cite{degennes1980weak,brand1989electromechanical,brand1994electrohydrodynamics,brand2006selected}, the rheology of smectic liquid crystals \cite{auernhammer2000undulation,auernhammer2002shear,auernhammer2005erratum}, uniaxial magnetic gels \cite{bohlius2004macroscopic,menzel2014bridging}, ferronematic and ferrocholesteric elastomers \cite{brand2014macroscopic,brand2015macroscopic}, and the behavior of active components in a gel-like environment \cite{brand2011macroscopic}. Relative translations and rotations have their meaning also in the present context, so we will include them into our approach. 

We proceed in the following way. First, in Sec.~\ref{sec:variables}, we introduce the set of variables that we use specifically to characterize the state of (visco)elastic composite materials. The resulting thermodynamic relations are presented in Sec.~\ref{sec:thermodynamics}. Next, we set up the static part of our macroscopic approach in Sec.~\ref{sec:statics}. After that, in Sec.~\ref{sec:dynamics}, the dynamic equations are derived. Then, in Sec.~\ref{sec:relstrains}, we demonstrate that the hydrodynamic-like theory is in accord with introducing and systematically using relative strains as a new macroscopic variable. For illustration, we address some reduced minimum examples of applying the theory in Sec.~\ref{sec:examples}, where also limitations of our approach are addressed. We insert a discussion in Sec.~\ref{sec:discussion}, before we conclude in Sec.~\ref{sec:conclusions}.

\section{Macroscopic variables} \label{sec:variables}

When setting up the theory, we first encounter the hydrodynamic variables already familiar from the macroscopic description of simple fluids \cite{landau1987fluid}. We will include them below. In the following, we first outline additional, possibly slowly relaxing variables characteristic for (visco)elastic composite materials. 

\begin{figure}
\centerline{\includegraphics[width=6.5cm]{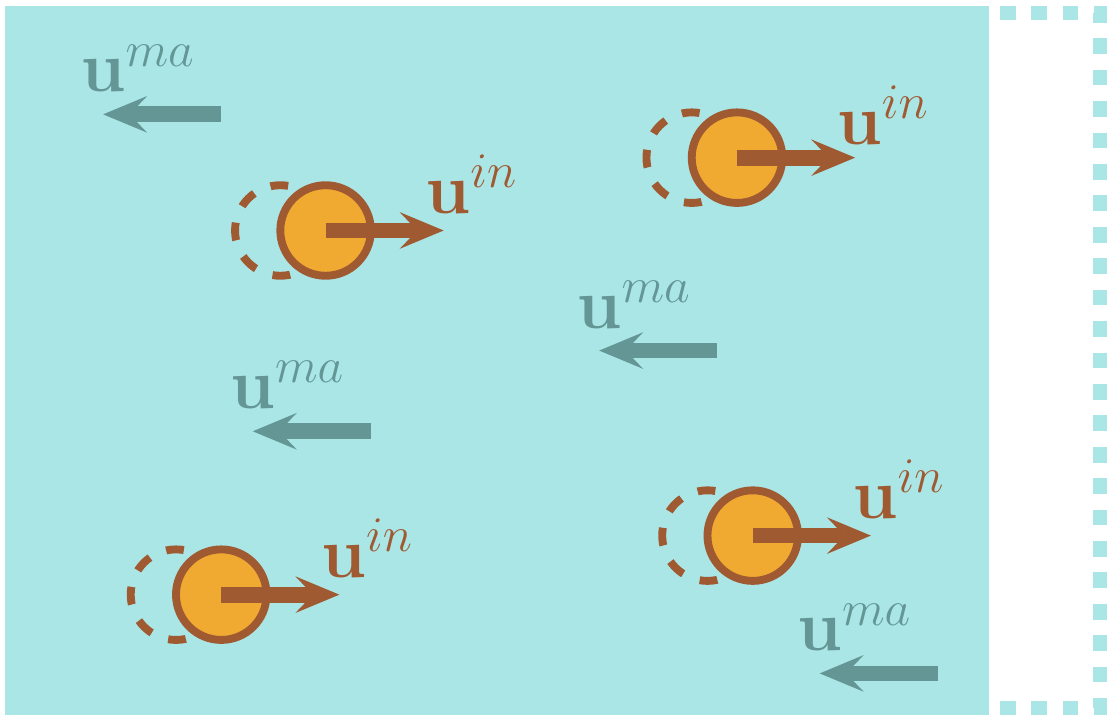}}
\caption{Illustration of relative translations. Dotted lines mark the initial state of the system. When the surrounding matrix is displaced in a way $\mathbf{u}^{ma}$ different from the displacements of the inclusions $\mathbf{u}^{in}$, a relative translation $\mathbf{u}^{rel}=\mathbf{u}^{ma}-\mathbf{u}^{in}$ results. 
}
\label{fig:reltrans}
\end{figure}

A relatively intuitive variable are \textit{relative translations}, see Fig.~\ref{fig:reltrans}. 
Typically, in linear elasticity theory, distortions of a material are described using the displacement field $\mathbf{u}(\mathbf{r},t)$ for the volume elements at positions $\mathbf{r}$. Here, we may introduce a macroscopic displacement field to characterize local displacements of the matrix, $\mathbf{u}^{ma}(\mathbf{r},t)$. However, the inclusions for some reason (see, e.g., the discussion in Sec.~\ref{sec:discussion}) may be displaced in a different way, captured by a different field $\mathbf{u}^{in}(\mathbf{r},t)$. If $\mathbf{u}^{ma}(\mathbf{r},t)\neq\mathbf{0}$ and $\mathbf{u}^{in}(\mathbf{r},t)\neq\mathbf{0}$, but both components are displaced in the same way, energetic contributions do not arise. In contrast to that, local differences between both displacements, i.e., relative translations $\mathbf{u}^{rel}(\mathbf{r},t)=\mathbf{u}^{ma}(\mathbf{r},t)-\mathbf{u}^{in}(\mathbf{r},t)$, typically cost energy. For instance, the matrix material around the inclusions is distorted when making place for relative translations of the inclusions. If relaxation processes of such relative translations are slow, it is justified to include them as macroscopic variables into our dynamic description. It should be noted that $\mathbf{u}^{in}(\mathbf{r},t)$ must be perceived as a local average over all inclusions within the volume element at position $\mathbf{r}$. In the following, the positional dependence will not be made explicit any more. 

\begin{figure}
\centerline{\includegraphics[width=7.8cm]{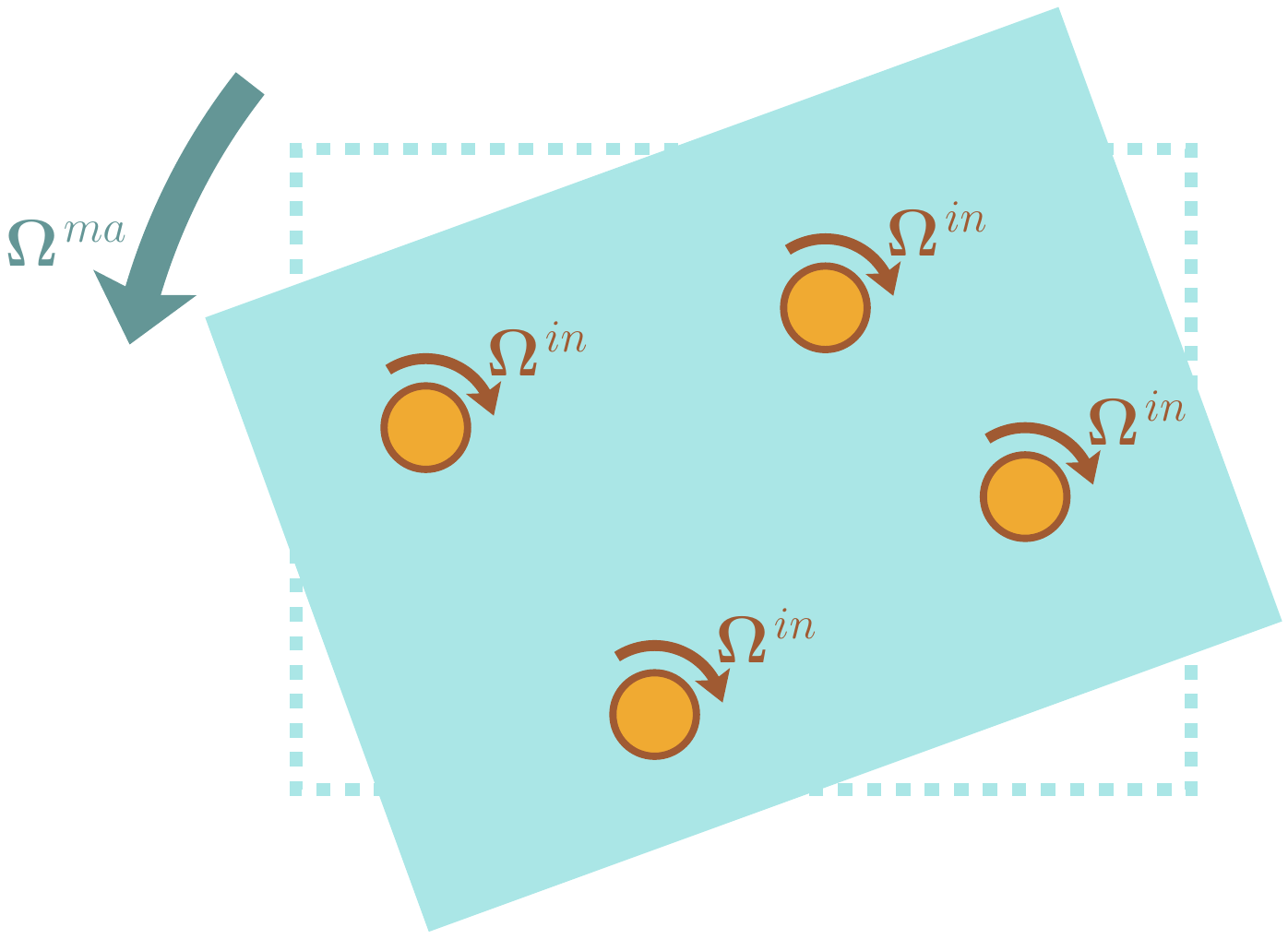}}
\caption{Illustration of relative rotations. Again, dotted lines mark the initial state. Rotating the matrix in a way $\mathbf{\Omega}^{ma}$ different from the rotations of the inclusions $\mathbf{\Omega}^{in}$ leads to relative rotations $\mathbf{\Omega}^{rel}=\mathbf{\Omega}^{ma}-\mathbf{\Omega}^{in}$. 
}
\label{fig:relrot}
\end{figure}

A similar situation arises for \textit{relative rotations}, as depicted in Fig.~\ref{fig:relrot}. 
Rotations can be described by antisymmetric tensors that contain the information on the rotation axis and magnitude. 
We express local matrix rotations by an antisymmetric tensor $\mathbf{\Omega}^{ma}$ and rotations of the inclusions by an antisymmetric tensor $\mathbf{\Omega}^{in}$. Here, the same reasoning as for relative translations arises. If the inclusions are rotated in a way different from their surrounding environment, relative rotations $\mathbf{\Omega}^{rel}=\mathbf{\Omega}^{ma}-\mathbf{\Omega}^{in}$ arise. If, for instance, matrix material is adsorbed on or linked to the inclusion surfaces, this leads to distortions of the matrix surrounding the inclusions and thus costs energy. Again, if corresponding relaxation processes are slow, relative rotations ought to be included as a macroscopic variable in a dynamic description. 

We remark an additional conceptual point. Local rotations of the matrix can in principle be expressed by the local matrix displacement field, i.e.\ $\mathbf{\Omega}^{ma}=[\nabla\mathbf{u}^{ma}-(\nabla\mathbf{u}^{ma})^T]/2$, where the superscript $^T$ marks the transpose. In contrast to that, $\mathbf{\Omega}^{in}$ cannot in general be obtained from an according macroscopic displacement field. This is because each inclusion may rotate around its individual axis, instead of all inclusions within a volume element rotating together around a common axis in a rigid-body configuration. Therefore, $\mathbf{\Omega}^{in}$ is obtained by averaging all rotation tensors for all inclusions within a local volume element of the material. 

Finally, we come to the strain variables. Here, a conceptual difference arises when compared to the two previous situations. (Quasi-)static translations and rotations can only contribute to the energy of the system when different components are translated or rotated relatively to each other. This is not the case for strain deformations. Straining one individual component by itself in general already costs energy. If the relaxation of the strain is slow enough, a corresponding macroscopic variable ought to be included into our dynamic description. 

\begin{figure}
\centerline{\includegraphics[width=4.5cm]{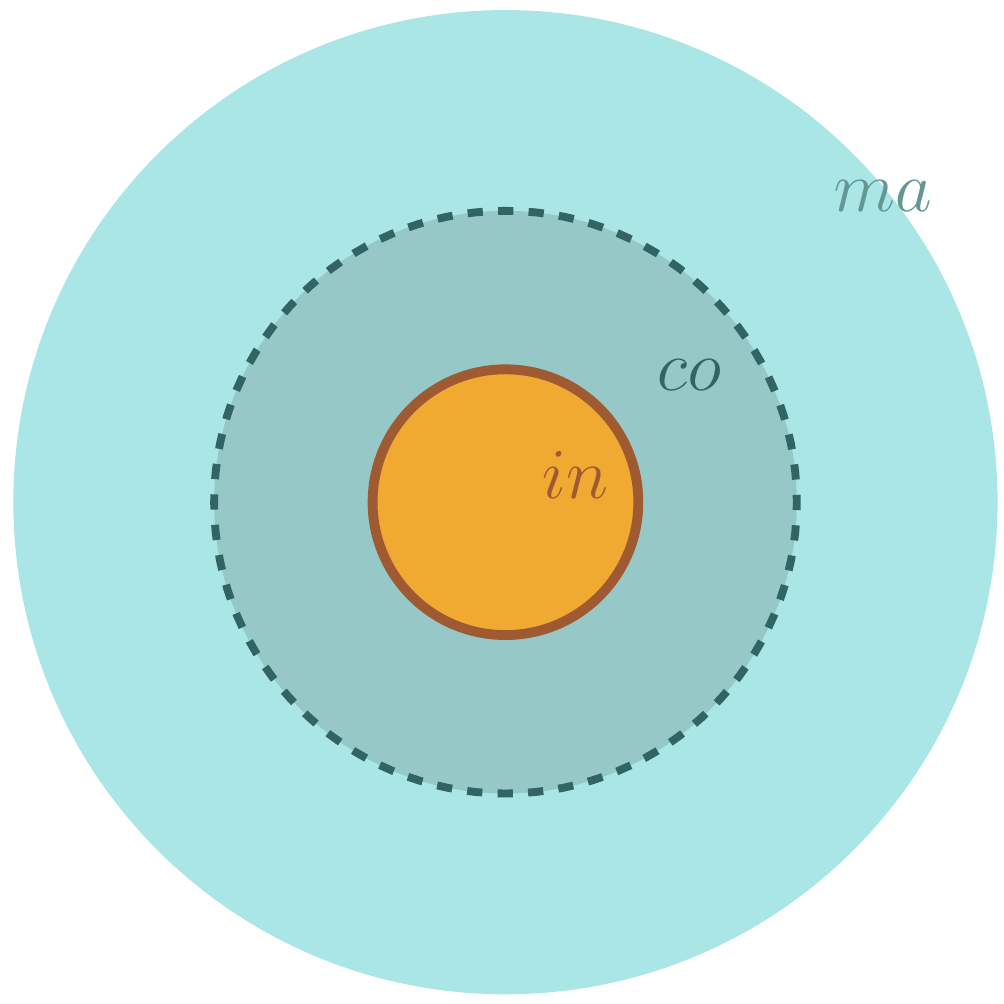}}
\caption{Illustration of the three zones that we distinguish in view of strain deformations: the embedding matrix material ($ma$), coupling zones ($co$) that directly surround the inclusions and couple them to the matrix, as well as the inclusions ($in$) themselves. 
}
\label{fig:threezones}
\end{figure}

As motivated in Sec.~\ref{sec:introduction}, we distinguish between three different zones within the materials, see Fig.~\ref{fig:threezones}. 
Naturally, the inclusions ($in$) are embedded in the surrounding matrix ($ma$). In addition to that, we resolve the effect of an explicit coupling zone ($co$) connecting these two components. These coupling zones may be intentionally fabricated \cite{yow2006formation,rauh2016seeded} or they may arise during the manufacturing process if the matrix material on the inclusion surfaces behaves differently from the bulk of the matrix \cite{landis2009explicit,gundermann2014investigation,huang2016buckling}. 

\begin{figure}
\centerline{\includegraphics[width=6.3cm]{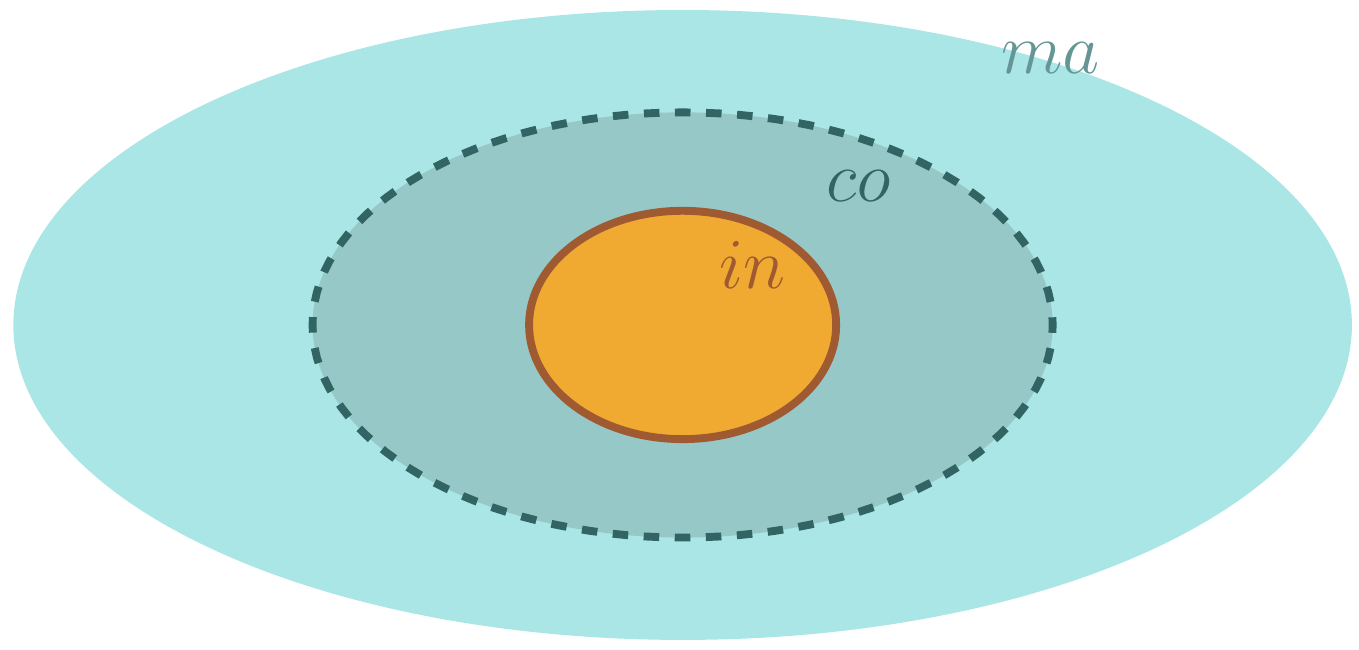}}
\caption{Illustration of the three zones $ma$, $co$, and $in$ in a strained state. As depicted here, the three zones may deform in different ways. However, the deformation of each zone by itself already costs energy. Therefore, three strain variables $\mathbf{U}^{ma}$, $\mathbf{U}^{co}$, and $\mathbf{U}^{in}$ are introduced. 
}
\label{fig:threezones_strained}
\end{figure}

When straining the material, e.g.\ when pulling on the matrix from outside, the different components may be strained in different ways, see Fig.~\ref{fig:threezones_strained}. 
This is most conceivable when we think of rigid inclusions that remain basically unstrained and then pull on the matrix from outside. Then the coupling zone must deform in yet another way to accommodate the presence of the non-deformable inclusions within the macroscopically deformed matrix. 
From this consideration, it also becomes clear that the coupling zones do not necessarily need to consist of a different kind of material. Instead, they can equally well serve as a simple concept to include strain inhomogeneities on the mesoscopic level into the overall macroscopic description, if necessary. 

\begin{figure}
\centerline{\includegraphics[width=8.5cm]{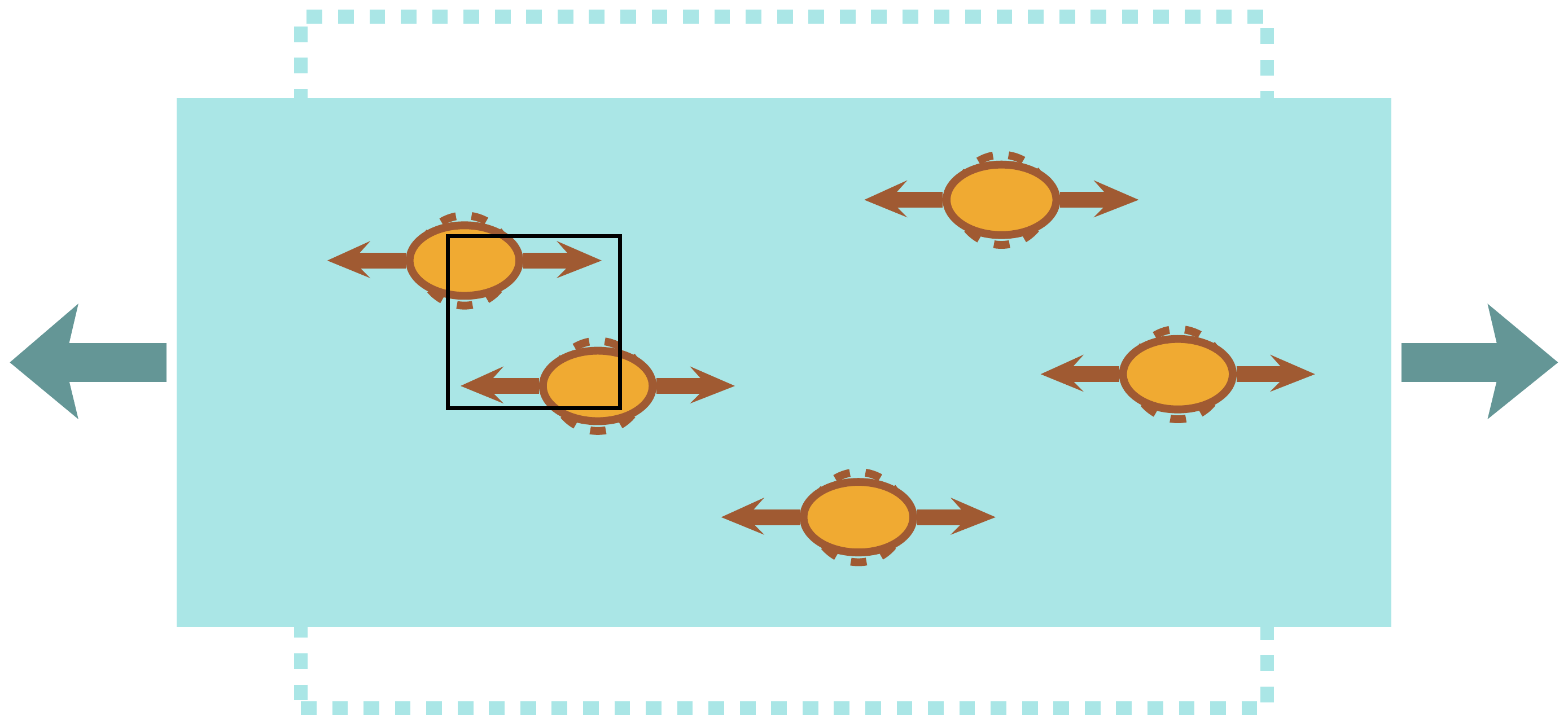}}
\caption{The deformation of each inclusion can be described by an individual strain tensor. Then, the macroscopic variable $\mathbf{U}^{in}$ must be obtained by averaging over all individual strain tensors for inclusions located within a corresponding volume element. In contrast to that, first averaging mesoscopic displacement fields for the inclusions leads to erroneous results. For instance, that part of the mesoscopic displacement field describing the strain deformations (indicated by the arrows) averaged over only one inclusion vanishes for symmetric strains. The same may apply by averaging over different inclusions, for instance those parts located within the black box.  
}
\label{fig:no_inclusion_displfield}
\end{figure}

As a consequence, we include three separate symmetric strain tensors $\mathbf{U}^{ma}$, $\mathbf{U}^{co}$, and $\mathbf{U}^{in}$ into our macroscopic theory. The matrix strain tensor may possibly be derived from the matrix displacement field $\mathbf{U}^{ma}=[\nabla\mathbf{u}^{ma}+(\nabla\mathbf{u}^{ma})^T]/2$ in the linear regime. However, we remark that for polymers \cite{pleiner1992macroscopic,muller2016transients,muller2016transientl} and viscoelastic materials in general \cite{temmen2000convective,pleiner2000structure} this approach may already be questioned. In the present case, 
this approach is not appropriate any more for our coupling zones and inclusions, if they are mesoscopically localized. 
One would, in that approach, first locally average the displacements to obtain the displacement field and then, from that displacement field, derive the strain tensor. However, 
for instance during a symmetric strain deformation,  
the displacements already vanish when averaged over one single mesoscopic inclusion, see further Fig.~\ref{fig:no_inclusion_displfield}. 
%
Therefore, during a coarse-graining procedure underlying our macroscopic description, it is not mesoscopic displacement fields but directly the strains of the inclusions that must be averaged. The macroscopic variable $\mathbf{U}^{in}$ 
must be obtained by averaging over the strain tensors for all inclusions within the corresponding volume element. 
The same applies for the coupling zones. 

\revision{At first glance, one might wonder, whether a description in terms of an additional strain tensor for the coupling zone is really necessary, when relative translations are already taken into account. However, it is readily seen that qualitatively different situations are addressed by these different variables. Consider, for instance, in Fig.~\ref{fig:threezones} the rigid inclusion shifted to the left by some external force, while the matrix is kept at rest. This corresponds to a relative translation. Then the coupling zone needs to be compressed left to the inclusion, while it gets expanded on the right side when the inclusion pulls on it. In contrast to that, the symmetric macroscopic strain tensors describe symmetric situations, as depicted in Fig.~\ref{fig:threezones_strained}. There, such left-right asymmetries cannot occur. The whole coupling zone can either be stretched or dilated along the horizontal principal axes. In other words, the macroscopic strain variables do not resolve locally inhomogeneous deformations within a single inclusion or coupling zone.}

In summary, we will work with the three macroscopic strain variables $\mathbf{U}^{ma}$, $\mathbf{U}^{co}$, and $\mathbf{U}^{in}$. As we will demonstrate in Sec.~\ref{sec:relstrains}, instead of explicitly working with different strain tensors, it is possible to use only one absolute strain variable, e.g.\ $\mathbf{U}^{ma}$, and then introduce \textit{relative strains} as macroscopic variables. Such an approach supplements the concepts of relative translations and relative rotations outlined above.

\section{Thermodynamic relations} \label{sec:thermodynamics}

As a first step, we specify the functional dependence of the total energy $E$ of the system, which needs to be an extensive variable \cite{toda1983statistical,pleiner2004general}: 
\begin{eqnarray}
  E &=& E(M, V, \mathbf{G}, S, Mc, M\mathbf{U}^{ma}, M\mathbf{U}^{co}, M\mathbf{U}^{in},\nonumber\\
  &&\qquad M\mathbf{u}^{rel}, M\mathbf{\Omega}^{rel}). 
  \label{eq:E}
\end{eqnarray}
Here, $M$ marks the total mass, $V$ the total volume, $\mathbf{G}$ the total momentum, and $S$ the total entropy of the system, while $c$ sets the concentration of inclusions. The other variables were introduced in Sec.~\ref{sec:variables}. 

To connect to a local hydrodynamic field description, the energy density $\epsilon=E/V$, mass density $\rho=M/V$, momentum density $\mathbf{g}=\mathbf{G}/V$, and entropy density $\sigma=S/V$ are introduced. Conjugate variables are defined as
\begin{eqnarray}
  \label{eq:mu}
  \mu &=& \frac{\partial E}{\partial M}, \\
  p &=& {}-\frac{\partial E}{\partial V}, \\
  {v}_i &=& \frac{\partial E}{\partial {G}_i}, \\
  T &=& \frac{\partial E}{\partial S}, \\
  \mu_c' &=& \frac{\partial E}{\partial (Mc)}, \\
  {{\Phi}^{x}_{ij}}' &=& \frac{\partial E}{\partial (M{U}^{x}_{ij})}, \\
  {{w}^{rel}_i}' &=& \frac{\partial E}{\partial (M{u}^{rel}_i)}, 
\end{eqnarray}
\begin{eqnarray}
  {{W}^{rel}_{ij}}' &=&  {}-\frac{\partial E}{\partial (M{\Omega}^{rel}_{ij})},
  \label{eq:W}
\end{eqnarray}
with $\mu$ the chemical potential; $p$ the pressure; $\mathbf{v}$ the velocity; $T$ the temperature; $\mu_c'$ a relative chemical potential for the inclusions; 
${\mathbf{\Phi}^{x}}'$ elastic stresses, where $x\in\{ma,co,in\}$; 
${\mathbf{w}^{rel}}'$ and ${\mathbf{W}^{rel}}'$ the conjugates to the relative translations and relative rotations, respectively. Since $\mathbf{U}^x$ is symmetric, so must be ${\mathbf{\Phi}^x}'$ ($x\in\{ma,co,in\}$). Likewise, since $\mathbf{\Omega}^{rel}$ is antisymmetric, so must be ${\mathbf{W}^{rel}}'$. 

Next, we calculate the differential of $E=\epsilon V$ using Eqs.~(\ref{eq:E})--(\ref{eq:W}). Collecting, on the one hand, terms in $\mathrm{d}V$ and, on the other hand, terms in $V$, we obtain
\begin{equation}
  \epsilon = \bar{\mu}\rho-p+\mathbf{v}\cdot\mathbf{g}+T\sigma
  \label{eq:epsilon}
\end{equation}
and the Gibbs relation
\begin{eqnarray}
  \mathrm{d}\epsilon &=& 
    \bar{\mu}\mathrm{d}\rho + \mathbf{v}\cdot\mathrm{d}\mathbf{g} 
      + T\mathrm{d}\sigma + \mu_c\mathrm{d}c 
      +\sum_{x} \mathbf{\Phi}^x : \mathrm{d}\mathbf{U}^x \nonumber\\
    &&{}
      +\mathbf{w}^{rel}\cdot\mathrm{d}\mathbf{u}^{rel}
      +\mathbf{W}^{rel}:\mathrm{d}\mathbf{\Omega}^{rel},
  \label{eq:depsilon}
\end{eqnarray}
where $x\in\{ma,co,in\}$. 
Here, we have defined
\begin{eqnarray}
  \label{eq:conjugate1}
  \bar{\mu} &=& \mu + \mu_c'c + \sum_x{\mathbf{\Phi}^x}':\mathbf{U}^x 
    \nonumber\\
    &&{}+ {\mathbf{w}^{rel}}'\cdot\mathbf{u}^{rel} 
    + {\mathbf{W}^{rel}}':\mathbf{\Omega}^{rel}
\end{eqnarray}
and
\begin{eqnarray}
  \mu_c &=& \mu_c'\rho,\\
  \mathbf{\Phi}^x &=& {\mathbf{\Phi}^x}'\rho,\\
  \mathbf{w}^{rel} &=& {\mathbf{w}^{rel}}'\rho,\\
  \mathbf{W}^{rel} &=& {\mathbf{W}^{rel}}'\rho,
  \label{eq:conjugate-last}
\end{eqnarray}
with $x\in\{ma,co,in\}$. 
Similarly, combining the above relations, we find for the differential of the pressure
\begin{eqnarray}
  \mathrm{d}p &=& \rho\mathrm{d}\bar{\mu} + \mathbf{g}\cdot\mathrm{d}\mathbf{v}
    +\sigma\mathrm{d}T-\mu_c\mathrm{d}c
    -\sum_x\mathbf{\Phi}^x:\mathrm{d}\mathbf{U}^x \nonumber\\
    &&{}
    -\mathbf{w}^{rel}\cdot\mathrm{d}\mathbf{u}^{rel}
    -\mathbf{W}^{rel}:\mathrm{d}\mathbf{\Omega}^{rel}.
\end{eqnarray}

Finally, we note that $\epsilon$ is a scalar and therefore must be invariant under rigid rotations of the whole system. From Eq.~(\ref{eq:depsilon}), this leads to the condition
\begin{eqnarray}
\lefteqn{2\sum_xU^x_{ik}\Phi^x_{kj} + u^{rel}_iw^{rel}_j + 2\Omega^{rel}_{ik}W^{rel}_{kj}} \nonumber\\
&=& 
2\sum_xU^x_{jk}\Phi^x_{ki} + u^{rel}_jw^{rel}_i + 2\Omega^{rel}_{jk}W^{rel}_{ki},
\label{eq:symmetrize_stress}
\end{eqnarray}
where summation over repeated indices is implied throughout. 
We will make use of this relation later to symmetrize the stress tensor. 

The basic assumption of hydrodynamics is that the above thermodynamic relations still apply locally at each space and time point $(\mathbf{r},t)$ when the system is moderately driven out of equilibrium \cite{kadanoff1963hydrodynamic,pleiner1996pattern,zwanzig2001nonequilibrium}.

\section{Statics} \label{sec:statics}

To establish the static part of the theory, we first derive an expression for the energy density $\epsilon$. For this purpose, we combine our macroscopic variables by symmetry arguments to corresponding contributions, 
\begin{eqnarray}
\epsilon &=& 
  \frac{1}{2\rho^2\kappa_{\rho}}(\delta\rho)^2 + \frac{T}{2C_V}(\delta\sigma)^2
  + \frac{\gamma}{2}(\delta c)^2 \nonumber\\
  &&{}
  + \frac{1}{\rho\alpha_S}(\delta\rho)(\delta\sigma)
  + \beta_{\rho}(\delta\rho)(\delta c) + \beta_{\sigma}(\delta\sigma)(\delta c) 
  \nonumber\\
  &&{}
  + \frac{1}{2\rho}g_ig_i \nonumber\\
  &&{}
  + \sum_x\left[ \chi^{x,\rho}_{ij}U^x_{ij}(\delta\rho)
      +\chi^{x,\sigma}_{ij}U^x_{ij}(\delta\sigma)
      +\chi^{x,c}_{ij}U^x_{ij}(\delta c) \right]\nonumber\\
  &&{}
  + \frac{1}{2}\sum_xc^x_{ijkl}U^x_{ij}U^x_{kl}
  + \frac{1}{2}\sum_{x,y;{x\neq y}} c^{x,y}_{ijkl}U^x_{ij}U^y_{kl}\nonumber\\
  &&{}
  + \frac{1}{2}d_{ij}u^{rel}_iu^{rel}_j 
  + \frac{1}{2}D_{ijkl}\Omega^{rel}_{ij}\Omega^{rel}_{kl},
\label{eq:expanded_epsilon}
\end{eqnarray}
where we have implicitly defined $c^{y,x}_{ijkl}=c^{x,y}_{ijkl}$ ($x,y\in\{ma,co,in\}$). 

Here, we only consider terms up to quadratic order, in accord with our overall scope of a linearized dynamic description. The parameter $\kappa_{\rho}$ denotes the compressibility, $C_V$ the specific heat, $\alpha_S$ a volume expansion coefficient, while $\gamma$, $\beta_{\rho}$, and $\beta_{\sigma}$ are similar coefficients related to changes in concentration \cite{pleiner1996pattern}.  
Next, the coefficients $\chi^{x,\rho}_{ij}$, $\chi^{x,\sigma}_{ij}$, and $\chi^{x,c}_{ij}$ describe the coupling of the strains of the different components to variations in density, entropy, and concentration, respectively. Deformation energies of the different components are characterized by $c^x_{ijkl}$ \cite{landau1986elasticity}, where here couplings between the strains of different components are possible as given by $c^{x,y}_{ijkl}$. Finally, $d_{ij}$ and $D_{ijkl}$ quantify the energetic contributions of relative translations and relative rotations, respectively. 

Confining ourselves for simplicity and brevity to isotropic systems in the present approach, we expand the listed material tensors as
\begin{eqnarray}
\chi^{x,\rho}_{ij} &=& \chi^{x,\rho} \delta_{ij}, \label{eq_chixrho} \\
\chi^{x,\sigma}_{ij} &=& \chi^{x,\sigma} \delta_{ij}, \\
\chi^{x,c}_{ij} &=& \chi^{x,c} \delta_{ij}, \label{eq_chixc} \\
c^{x}_{ijkl} &=& \frac{1}{2}c_1^x
  \left(\delta_{ik}\delta_{jl}+\delta_{il}\delta_{jk}\right)
  +c_2^x\delta_{ij}\delta_{kl}, 
\label{eq_cx}
\\
c^{x,y}_{ijkl} &=& \frac{1}{2}c_1^{x,y}
  \left(\delta_{ik}\delta_{jl}+\delta_{il}\delta_{jk}\right)
  +c_2^{x,y}\delta_{ij}\delta_{kl}, 
\label{eq_cxy}
\\
d_{ij} &=& d \delta_{ij}, \\
D_{ijkl} &=& \frac{1}{2}D\epsilon_{rij}\epsilon_{rkl}
  = \frac{1}{2}D\left(\delta_{ik}\delta_{jl}-\delta_{il}\delta_{jk}\right),
\end{eqnarray}
where $\delta_{ij}$ marks the Kronecker delta and $\epsilon_{ijk}$ the Levi-Civita tensor. 

In principle, also terms coupling $\mathbf{\Omega}^{rel}$ to $\delta\rho$, $\delta c$, $\delta\sigma$, and $\mathbf{U}^x$ ($x\in\{ma,co,in\}$) are allowed by symmetry. However, since $\mathbf{\Omega}^{rel}$ is antisymmetric (and $\mathbf{U}^x$ is symmetric), they vanish. 

Following Eqs.~(\ref{eq:depsilon}) and (\ref{eq:expanded_epsilon}), we find for the conjugate variables 
\begin{eqnarray}
\bar{\mu} = \frac{\partial\epsilon}{\partial\rho} 
  &=& \frac{1}{\rho^2\kappa_{\rho}}(\delta\rho) 
      + \frac{1}{\rho\alpha_S}(\delta\sigma)
      + \beta_{\rho}(\delta c) \nonumber\\
  &&{}+ \sum_x\chi^{x,\rho}U^x_{ii},  \\
T = \frac{\partial\epsilon}{\partial\sigma} 
  &=& \frac{T}{C_V}(\delta\sigma) + \frac{1}{\rho\alpha_S}(\delta\rho)
      + \beta_{\sigma}(\delta c) \nonumber\\
  &&{}+ \sum_x\chi^{x,\sigma}U^x_{ii},  \\
{\mu}_c = \frac{\partial\epsilon}{\partial c} 
  &=& \gamma(\delta c) + \beta_{\rho}(\delta\rho) 
      + \beta_{\sigma}(\delta\sigma)
       \nonumber\\
  &&{}+ \sum_x\chi^{x,c}U^x_{ii},  \\
v_i = \frac{\partial\epsilon}{\partial g_i} &=& \frac{1}{\rho}g_i,  \\ 
\Phi^x_{ij} = \frac{\partial\epsilon}{\partial U^x_{ij}} 
  &=& c^x_{ijkl}U^x_{kl} + \sum_{y;y\neq x}c^{x,y}_{ijkl}U^y_{kl} \nonumber\\
  &&{}
      + \chi^{x,\rho}\delta_{ij}(\delta\rho)
      + \chi^{x,\sigma}\delta_{ij}(\delta\sigma)\nonumber\\
  &&{}
      + \chi^{x, c}\delta_{ij}(\delta c),  
\label{eq_Phi}\\ 
w^{rel}_i = \frac{\partial\epsilon}{\partial u^{rel}_i} &=& d\,u^{rel}_i,  \\ 
W^{rel}_{ij} = {}-\frac{\partial\epsilon}{\partial \Omega^{rel}_{ij}} 
  &=& {}-D\,\Omega^{rel}_{ij},  
\end{eqnarray}
where not all tensors have been expanded in favor of brevity.

\section{Dynamics} \label{sec:dynamics}

In the next step, we turn to the dynamics of the system. For this purpose, currents $\mathbf{g}=\rho\mathbf{v}$, $\mathbf{j}^{\sigma}$, $\mathbf{j}^c$, $\bm{\sigma}$, and $\mathbf{j}^{\epsilon}$ are introduced in the context of conservation laws and quasi-currents $\mathbf{Y}^x$ ($x\in\{ma,co,in\}$), $\mathbf{q}$, and $\mathbf{Q}$ in the remaining cases: 
\begin{eqnarray}
  \partial_t\rho + \nabla_i\rho v_i &=& 0, \\
  \partial_t\sigma + \nabla_i\sigma v_i + \nabla_i j^{\sigma}_i &=& \frac{R}{T}, \\
  \partial_t c + v_i\nabla_i c + \nabla_i j^c_i &=& 0, \\
  \partial_t g_i + \nabla_jg_iv_j + \nabla_j\sigma_{ij} &=& 0, 
\label{eq_dyng}\\
  \partial_t \epsilon + \nabla_i\left[(\epsilon+p)v_i\right] 
    + \nabla_ij^{\epsilon}_i &=& 0,\\
  \partial_t U^x_{ij} + v_k\nabla_k U^x_{ij} + Y^x_{ij} &=& 0, 
\label{eq_dynU}\\
  \partial_t u^{rel}_i + v_j\nabla_j u^{rel}_i + q_i &=& 0, \\
  \partial_t \Omega^{rel}_{ij} + v_k\nabla_k\Omega^{rel}_{ij} + Q_{ij} &=& 0. 
\end{eqnarray}
$R$ denotes the entropy production rate, where we require $R=0$ for reversible and $R>0$ for irreversible processes. 
Since the energy density is related to the other variables by the Gibbs relation Eq.~(\ref{eq:depsilon}), we do not need to explicitly keep track of the energy conservation law in the following \cite{pleiner2004general}. 

Inserting these equations into the Gibbs relation Eq.~(\ref{eq:depsilon}), we find 
\begin{eqnarray}
R &=& {}-j^{\sigma}_i\nabla_i T - j^c_i\nabla_i\mu_c -\sigma'_{ij}\nabla_jv_i
      \nonumber\\ 
      &&{}
      +\sum_x \Phi^x_{ij}Y^x_{ji} + w^{rel}_iq_i + W^{rel}_{ij}Q_{ji},
  \label{eq:R}
\end{eqnarray}
except for a divergence term that does not need to be considered in the following \cite{pleiner2004general}. 
In this context, using Eq.~(\ref{eq:epsilon}), we have defined
\begin{equation}\label{eq_sigma-sigmaprime-p}
\sigma_{ij} = \sigma'_{ij} + p\delta_{ij}.
\end{equation}
Eq.~(\ref{eq:R}) identifies the generalized forces $\nabla T$, $\nabla\mu_c$, $\nabla\mathbf{v}$, $\mathbf{\Phi}^x$ ($x\in\{ma,co,in\}$), $\mathbf{w}^{rel}$, and $\mathbf{W}^{rel}$.  

Next, we split our currents into reversible and dissipative parts, 
\begin{eqnarray}
j^{\sigma}_i &=& j^{\sigma,rev}_i + j^{\sigma,dis}_i, \\
j^c_i &=& j^{c,rev}_i + j^{c,dis}_i, \\
\sigma'_{ij} &=& \sigma^{\prime\: rev}_{ij} + \sigma^{\prime\: dis}_{ij}, \label{eq_sigmasplit}\\
Y^x_{ij} &=& Y^{x,rev}_{ij} + Y^{x,dis}_{ij},
\label{eq_Ysplit} \\
q_i &=& q^{rev}_i + q^{dis}_i, \\
Q_{ij} &=& Q^{rev}_{ij} + Q^{dis}_{ij},
\end{eqnarray}
where the reversible currents need to respect $R=0$ when inserted into Eq.~(\ref{eq:R}), while the dissipative parts must satisfy $R\geq0$.

\subsection{Reversible dynamics}

The reversible parts of the currents are constructed by symmetry arguments from the generalized forces. For macroscopic variables that are even (odd) under time reversal, the corresponding reversible currents must be odd (even) \cite{pleiner1996pattern}. 

We first list our results and then add some explanations: 
\begin{eqnarray}
j^{\sigma,rev}_i &=& 0, \\[.1cm]
j^{c,rev}_i &=& 0, \\[.1cm]
\sigma^{\prime\: rev}_{ij} &=& a\left[ 2\sum_x U^x_{ik}\Phi^x_{kj} 
                                       + u^{rel}_iw^{rel}_j
                                       + 2\Omega^{rel}_{ik}W^{rel}_{kj} \right]\qquad
  \nonumber\\[.1cm]
  && {}-\sum_x b^x \Phi_{ij}^x,
\label{sigmarev} \\[.1cm]
Y^{x,rev}_{ij} &=& a\left[ U^x_{ik}(\nabla_jv_k) + U^x_{jk}(\nabla_iv_k) \right]
  -b^xA_{ij},\qquad
\label{eq_Yrev}
\\[.1cm]
q^{rev}_{i} &=&  a\,u^{rel}_j(\nabla_iv_j),
\label{qrev}
\\[.1cm]
Q^{rev}_{ij} &=& a\left[ \Omega^{rel}_{ik}(\nabla_j v_k) 
                       - \Omega^{rel}_{jk}(\nabla_i v_k) \right]. 
\label{Qrev}
\end{eqnarray}

In the derivation of the above expressions, we used that $\mathbf{U}^x$ and $\mathbf{\Phi}^x$ are symmetric ($x\in\{ma,co,in\}$), whereas $\mathbf{\Omega}^{rel}$ and $\mathbf{W}^{rel}$ are antisymmetric. Moreover, to maintain the symmetry of $\mathbf{U}^x$, only symmetrized terms may enter the right-hand side of Eq.~(\ref{eq_Yrev}). Therefore, we here need to include the symmetrized velocity gradient tensor $\mathbf{A}$, where 
\begin{equation}
A_{ij}=\left(\nabla_iv_j+\nabla_jv_i\right)/2. 
\label{eq_Aij}
\end{equation}
Similarly, the antisymmetry of $\mathbf{\Omega}^{rel}$ must be maintained. Therefore the right-hand side of Eq.~(\ref{Qrev}) must be antisymmetric and $A_{ij}$ may not enter. 

\revision{Our scope is to present the linearized part of the theory. Nevertheless, we have added some of the possible nonlinear couplings in Eqs.~(\ref{sigmarev})--(\ref{Qrev}). The nonlinear coupling terms $U_{ik}^x\Phi_{kj}^x$ in $\sigma^{\prime\: rev}_{ij}$ as well as $\left(U_{ik}^x\nabla_jv_k+U_{jk}^x\nabla_iv_k\right)$ in $Y_{ij}^{x,rev}$ ($x\in\{ma,co,in\}$) are included to connect with a previous systematic, symmetry-based, and generalized nonlinear approach to one-component viscoelastic systems \cite{temmen2000convective,pleiner2000structure,temmen2001temmen,pleiner2004nonlinear, muller2016transients,muller2016transientl}. There, they resulted from the requirement that the physics be independent of the orientation of the system (as long as external fields are absent or likewise reoriented) \cite{temmen2000convective,pleiner2000structure,pleiner2004nonlinear}. For $a=1$, the form $\partial_t U^x_{ij} + v_k\nabla_k U^x_{ij} + U^x_{ik}(\nabla_jv_k) + U^x_{jk}(\nabla_iv_k)$ resulting from Eqs.~(\ref{eq_dynU}) and (\ref{eq_Yrev}) is typically referred to as the lower convected derivative \cite{temmen2001temmen,pleiner2004nonlinear}.} 

\revision{In parallel to Eq.~(\ref{eq_Yrev}) and in view of angular momentum conservation, according nonlinear terms were likewise included in Eqs.~(\ref{qrev}) and (\ref{Qrev}). At first glance, these nonlinear contributions could enter with a different coupling coefficient in each equation. Here, we set them identical to satisfy the requirement that angular momentum be conserved. 
For this purpose, it must be possible to formulate the stress tensor in a symmetric form \cite{martin1972unified}. Eq.~(\ref{eq:symmetrize_stress}) is applied to this end. 
It requires the same coefficient, called $a$ in the above expressions, on all these terms.} After symmetrization, the reversible part of the stress tensor reads
\begin{eqnarray}
\sigma^{\prime\:rev}_{ij} &=& a\bigg[ \sum_x\left( U^x_{ik}\Phi^x_{kj}
  + U^x_{jk}\Phi^x_{ki} \right)  \nonumber\\
  &&{}
  \quad+\frac{1}{2}\left( u^{rel}_i w^{rel}_j + u^{rel}_j w^{rel}_i \right)  
  \nonumber\\
  &&{}
  \quad+ \Omega^{rel}_{ik}W^{rel}_{kj} + \Omega^{rel}_{jk}W^{rel}_{ki} \bigg] 
  \nonumber\\
  &&{} -\sum_x b^x\Phi^x_{ij}.
\label{eq_sigma-prime-rev}
\end{eqnarray}
\revision{In Refs.~\onlinecite{temmen2000convective} and \onlinecite{pleiner2000structure}, for a one-component system, it was demonstrated by explicit calculation that $a=1$ and the corresponding $b^x=1$, based on symmetry arguments. The calculation assumes that a displacement field exists. In our minimal examples below, we only consider the linear contributions. There we set all $b^x$ equal and scale them out or we set some of them to zero when at one point we decouple the system.}

We remark that an additional antisymmetric contribution $fR_{ij}$ in Eq.~(\ref{Qrev}), with $R_{ij}=\left(\nabla_iv_j-\nabla_jv_i\right)/2$, would maintain the antisymmetric nature of $\mathbf{\Omega}^{rel}$. However, this implies an antisymmetric contribution $fW_{ij}$ to $\sigma^{\prime\:rev}_{ij}$ that may enable changes in angular momentum. Such a situation appears plausible when, for instance, rotational torques are applied on anisotropic inclusions by external fields, see also the discussion in Sec.~\ref{sec:discussion}. If relative rotations cost energy, such torques will be transmitted to the whole material and induce rotational motion.

\subsection{Dissipative dynamics}

Finally, the dissipative parts of the currents are obtained by first expanding the entropy production rate $R$ into the generalized forces. Using symmetry arguments, we obtain
\begin{eqnarray}
2R &=& \kappa_{ij}(\nabla_i T)(\nabla_j T) + D_{ij}(\nabla_i\mu_c)(\nabla_j\mu_c)
       \nonumber\\
       &&{}
       +2D^{(T)}_{ij}(\nabla_i T)(\nabla_j\mu_c)
       \nonumber\\
       &&{}
       +\nu_{ijkl}(\nabla_i v_j)(\nabla_k v_l)
       \nonumber\\
       &&{}
       +\sum_x \zeta^x_{ijkl}\Phi^x_{ij}\Phi^x_{kl}
       +\sum_{x,y;x\neq y} \zeta^{x,y}_{ijkl}\Phi^x_{ij}\Phi^y_{kl}
       \nonumber\\
       &&{}
       +\xi^{rel}_{ij}w^{rel}_i w^{rel}_j 
       \nonumber\\
       &&{}
       +2\xi^{rel,T}_{ij}w^{rel}_i(\nabla_j T)
       +2\xi^{rel,c}_{ij}w^{rel}_i(\nabla_j \mu_c)
       \nonumber\\
       &&{}
       +\upsilon_{ijkl}W^{rel}_{ij} W^{rel}_{kl}.
  \label{eq:2R}
\end{eqnarray}
In this expression, the coefficients ${\kappa}_{ij}$ are related to heat conduction \cite{pleiner1996pattern}, ${D}_{ij}$ and $D^{(T)}_{ij}$ describe diffusion and thermodiffusion \cite{gutkowicz1979rayleigh,knobloch1980convection,brand1983convective, knobloch1986oscillatory,pleiner1996pattern,puljiz2016thermophoretically}, respectively, while $\nu_{ijkl}$ comprise the viscosities \cite{pleiner1996pattern}. Next, the coefficients $\zeta^x_{ijkl}$ include irreversible relaxation processes of strains \cite{pleiner2004general,pleiner2004generalaip}, e.g.\ due to disentanglement of intertwined high-molecular-weight polymer chains \cite{degennes1979scaling,doi2007theory}. In our three-component system, couplings between the stresses of the different components are possible, represented by the coefficients $\zeta^{x,y}_{ijkl}$. Dissipative relaxation of relative translations may occur as described by the coefficients $\xi^{rel}_{ij}$, where couplings to temperature and concentration gradients are possible as given by $\xi^{rel,T}_{ij}$ and $\xi^{rel,c}_{ij}$. Finally, potential dissipative relaxation of relative rotations is characterized by $\upsilon_{ijkl}$ in the present framework. 

As before, considering for simplicity an isotropic system, the coefficient tensors are expanded as
\begin{eqnarray}
\kappa_{ij} &=& \kappa\delta_{ij}, \\
D_{ij} &=& D\delta_{ij}, \\
D^{(T)}_{ij} &=& D^{(T)}\delta_{ij}, \\
\nu_{ijkl} &=& \frac{1}{2}\nu_1
  \left(\delta_{ik}\delta_{jl}+\delta_{il}\delta_{jk}\right)
  +\nu_2\delta_{ij}\delta_{kl}, 
\label{eq_nu}\\
\zeta^{x}_{ijkl} &=& \frac{1}{2}\zeta^x_1
  \left(\delta_{ik}\delta_{jl}+\delta_{il}\delta_{jk}\right)
  +\zeta^x_2\delta_{ij}\delta_{kl}, 
\label{eq_zetax}
\\
\zeta^{x,y}_{ijkl} &=& \frac{1}{2}\zeta^{x,y}_1
  \left(\delta_{ik}\delta_{jl}+\delta_{il}\delta_{jk}\right)
  +\zeta^{x,y}_2\delta_{ij}\delta_{kl}, 
\label{eq_zetaxy}\\
\xi^{rel}_{ij} &=& \xi^{rel}\delta_{ij}, 
\end{eqnarray}
\begin{eqnarray}
\xi^{rel,T}_{ij} &=& \xi^{rel,T}\delta_{ij}, \\
\xi^{rel,c}_{ij} &=& \xi^{rel,c}\delta_{ij}, \\
\upsilon_{ijkl} &=& \frac{1}{2}\upsilon\,\epsilon_{rij}\epsilon_{rkl}=\frac{1}{2}\upsilon\left(\delta_{ik}\delta_{jl}-\delta_{il}\delta_{jk}\right).
\end{eqnarray}

In principle, couplings between $\mathbf{W}^{rel}$ and $\mathbf{\Phi}^x$ ($x\in\{ma,co,in\}$) would be allowed by symmetry in Eq.~(\ref{eq:2R}). Yet, since $\mathbf{W}^{rel}$ is antisymmetric and the $\mathbf{\Phi}^x$ are symmetric, these terms vanish. 

Systematically taking derivatives of $R$ with respect to the generalized forces as prescribed by Eq.~(\ref{eq:R}), we find for the dissipative parts of the currents: 
\begin{eqnarray}
j^{\sigma,dis}_i &=& {}-\frac{\partial R}{\partial(\nabla_i T)} 
  \nonumber\\[.1cm]
  &=&
  {}-\kappa(\nabla_iT) - D^{(T)}(\nabla_i\mu_c) - \xi^{rel,T}w_i^{rel},\qquad
\\[.1cm]
j^{c,dis}_i &=& {}-\frac{\partial R}{\partial(\nabla_i \mu_c)} 
  \nonumber\\[.1cm]
  &=& 
  {}-D(\nabla_i\mu_c) - D^{(T)}(\nabla_iT) - \xi^{rel,c}w^{rel}_i,\qquad
\\[.1cm]
\sigma^{\prime\: dis}_{ij} &=& {}-\frac{\partial R}{\partial(\nabla_j v_i)} 
  \:=\: {}-\nu_{ijkl}(\nabla_kv_l),
\label{eq_sigmadis}
\\[.1cm]
Y^{x,dis}_{ij} &=& \frac{\partial R}{\partial \Phi_{ij}^x} 
  \nonumber\\[.1cm]
  &=& \zeta^x_{ijkl}\Phi_{kl}^x + \sum_{y; y\neq x}\zeta^{x,y}_{ijkl}\Phi_{kl}^y,
\label{eq_Ydis}
\\[.1cm]
q^{dis}_i &=& \frac{\partial R}{\partial w_i^{rel}} 
  \nonumber\\[.1cm]
  &=& \xi^{rel}w_i^{rel} + \xi^{rel,T}(\nabla_iT) + \xi^{rel,c}(\nabla_i\mu_c),
  \label{eq_qdis}
\\[.1cm]
Q^{dis}_{ij} &=& {}-\frac{\partial R}{\partial W_{ij}^{rel}} \:=\: {}-\upsilon\, W^{rel}_{ij}.
\end{eqnarray}
Again, in favor of brevity not all tensors have been expanded. 
This list of dissipative currents completes our macroscopic description.

\section{Relative strains} \label{sec:relstrains}

Let us now come back to our claim that the theory can be formulated using as macroscopic variables the \textit{relative strains} between the components. 
For brevity, let us here only consider a matrix component and inclusions, neglecting the influence of the coupling zones. That is, in the above description all variables carrying the superscript $^{co}$ are set to zero. 

We start from the pure strain part $\epsilon^{str}$ in the energy density $\epsilon$ in Eq.~(\ref{eq:expanded_epsilon}), 
\begin{equation}
\epsilon^{str}  =  \frac{1}{2}c_{ijkl}^{ma}U_{ij}^{ma}U_{kl}^{ma} 
                   + \frac{1}{2}c_{ijkl}^{in}U_{ij}^{in}U_{kl}^{in}
                   + c_{ijkl}^{ma,in}U_{ij}^{ma}U_{kl}^{in},
\end{equation}
and rewrite it as
\begin{eqnarray}
\tilde{\epsilon}^{str}
  &=& \frac{1}{2}c_{ijkl}U_{ij}^{ma}U_{kl}^{ma}
      + \frac{1}{2}D^{(1)}_{ijkl}U_{ij}^{rel}U_{kl}^{rel} \nonumber\\
  && {}    + D^{(2)}_{ijkl}U_{ij}^{ma}U_{kl}^{rel}. 
\label{eq_estr}
\end{eqnarray}
In this context we have defined the new variable of relative strains
\begin{equation}
\mathbf{U}^{rel} = \mathbf{U}^{ma}-\mathbf{U}^{in} 
\label{eq_Urel}
\end{equation}
as well as the coefficient tensors
\begin{eqnarray}
c_{ijkl} &=& c_{ijkl}^{ma} + 2c_{ijkl}^{ma,in} + c_{ijkl}^{in}, \\
D_{ijkl}^{(1)} &=& c_{ijkl}^{in}, \\
D_{ijkl}^{(2)} &=& {}-\left( c_{ijkl}^{ma,in} + c_{ijkl}^{in} \right).
\label{eq_D2}
\end{eqnarray}
From Eq.~(\ref{eq_estr}) we infer one important difference between the new macroscopic variables of relative strains and the previously introduced variables of relative translations \cite{brand1983hydrodynamics,pleiner1992macroscopic} and relative rotations \cite{degennes1980weak,brand1989electromechanical,brand1994electrohydrodynamics, brand2006selected,auernhammer2000undulation,auernhammer2002shear, auernhammer2005erratum,bohlius2004macroscopic,menzel2014bridging, brand2014macroscopic,brand2015macroscopic,brand2011macroscopic}. In the latter cases, \textit{only} the \textit{relative} translations and \textit{relative} rotations between the two components cost energy. However, when considering strains, already the deformation of \textit{one} of the components by itself contributes to the energy density. This is why, in addition to the relative strains between the components, we also find one \textit{absolute} strain variable in Eq.~(\ref{eq_estr}), here chosen to be $\mathbf{U}^{ma}$. 

Eq.~(\ref{eq_Urel}) sets the concept. The objective of the whole remaining section is to demonstrate that instead of initially using the absolute strains $\mathbf{U}^{ma}$ and $\mathbf{U}^{in}$ as macroscopic variables, one could have equally proceeded by building the theory on $\mathbf{U}^{ma}$ and $\mathbf{U}^{rel}$. For this purpose, we reformulate all remaining corresponding expressions accordingly. 

First, let us rewrite the couplings of $\mathbf{U}^{ma}$ and $\mathbf{U}^{in}$ to the scalars $\delta\rho$, $\delta\sigma$, and $\delta c$ in Eq.~(\ref{eq:expanded_epsilon}) as
\begin{equation}
\tilde{\epsilon}^{\chi} = \sum_{z\in\{\rho,\sigma,c\}} \left[
                          \tilde{\chi}_{ij}^{ma,z}U_{ij}^{ma}(\delta z)
                          + \tilde{\chi}_{ij}^{rel,z}U_{ij}^{rel}(\delta z) \right],
\label{eq_echi}
\end{equation}
where we have defined
\begin{eqnarray}
\tilde{\chi}_{ij}^{ma,z} &=& \chi_{ij}^{ma,z} + \chi_{ij}^{in,z}, 
\label{eq_chima}
\\
\tilde{\chi}_{ij}^{rel,z} &=& {}-\chi_{ij}^{in,z}, 
\label{eq_chirel}
\end{eqnarray}
with $z\in\{\rho,\sigma,c\}$. 

Turning now to the conjugate variables, we expand 
\begin{eqnarray}
\tilde{\chi}_{ij}^{ma,z} &=& \tilde{\chi}^{ma,z} \delta_{ij}, \\
\tilde{\chi}_{ij}^{rel,z} &=& \tilde{\chi}^{rel,z} \delta_{ij},
\end{eqnarray} 
in analogy to Eqs.~(\ref{eq_chixrho})--(\ref{eq_chixc}), where again $z\in\{\rho,\sigma,c\}$. As a consequence, the strain-dependent parts in $\bar{\mu}$, $T$, and $\mu_c$ now read 
\begin{eqnarray}
\bar{\mu}^{\chi,str} &=& \tilde{\chi}^{ma,\rho}U_{ii}^{ma}+\tilde{\chi}^{rel,\rho}U_{ii}^{rel},
\\
T^{\chi,str} &=& \tilde{\chi}^{ma,\sigma}U_{ii}^{ma}+\tilde{\chi}^{rel,\sigma}U_{ii}^{rel},
\\
\mu_c^{\chi,str} &=& \tilde{\chi}^{ma,c}U_{ii}^{ma}+\tilde{\chi}^{rel,c}U_{ii}^{rel}.
\end{eqnarray}

The conjugate variables to the absolute and relative strains $\mathbf{U}^{ma}$ and $\mathbf{U}^{rel}$, respectively, are now obtained from Eqs.~(\ref{eq_estr}) and (\ref{eq_echi}) as 
\begin{eqnarray}
\tilde{\Phi}^{ma}_{ij} &=& \frac{\partial\left(\tilde{\epsilon}^{str}+\tilde{\epsilon}^{\chi}\right)}{\partial U^{ma}_{ij}} 
\nonumber\\[.1cm]
  &=& c_{ijkl}U^{ma}_{kl} + D^{(2)}_{ijkl}U^{rel}_{kl} 
        + \sum_{z\in\{\rho,\sigma,c\}} \tilde{\chi}^{ma,z}_{ij}(\delta z) 
\nonumber\\[.1cm]
  &=& (c^{ma}_{ijkl}+c^{ma,in}_{ijkl})U^{ma}_{kl}
      + (c^{ma,in}_{ijkl}+c^{in}_{ijkl})U^{in}_{kl}
\nonumber\\[.1cm]
  && {}
      + \sum_{z\in\{\rho,\sigma,c\}} \left[ \chi^{ma,z}_{ij}(\delta z)
          + \chi^{in,z}_{ij}(\delta z) \right] 
\nonumber\\[.1cm]
  &=& \Phi^{ma}_{ij} + \Phi^{in}_{ij}, 
\\[.1cm]
\tilde{\Phi}^{rel}_{ij} &=& \frac{\partial\left(\tilde{\epsilon}^{str}+\tilde{\epsilon}^{\chi}\right)}{\partial U^{rel}_{ij}} 
\nonumber\\[.1cm]
  &=& D^{(1)}_{ijkl}U^{rel}_{kl} + D^{(2)}_{ijkl}U^{ma}_{kl} 
        + \sum_{z\in\{\rho,\sigma,c\}} \tilde{\chi}^{rel,z}_{ij}(\delta z) 
\nonumber\\[.1cm]
  &=& {}-c^{ma,in}_{ijkl}U^{ma}_{kl}
      -c^{in}_{ijkl}U^{in}_{kl}
      - \sum_{z\in\{\rho,\sigma,c\}} \chi^{in,z}_{ij}(\delta z)
\nonumber\\[.1cm]
  &=& {}-\Phi^{in}_{ij}, 
\end{eqnarray}
where we used Eqs.~(\ref{eq_Urel})--(\ref{eq_D2}), (\ref{eq_chima}), and (\ref{eq_chirel}) to connect to the previous expressions in Eq.~(\ref{eq_Phi}). 

The dynamics for $\mathbf{U}^{ma}$ and $\mathbf{U}^{rel}$ is directly obtained from Eqs.~(\ref{eq_dynU}) as the equation for $\mathbf{U}^{ma}$ and by subtracting from it the equation for $\mathbf{U}^{in}$, respectively:
\begin{eqnarray}
\partial_t U^{ma}_{ij} + v_k\nabla_k U^{ma}_{ij} + Y^{ma}_{ij} &=& 0, 
\label{eq_dynUma}
\\
\partial_t U^{rel}_{ij} + v_k\nabla_k U^{rel}_{ij} + Y^{rel}_{ij} &=& 0. 
\end{eqnarray} 
Here, we introduced 
\begin{equation}
\mathbf{Y}^{rel} = \mathbf{Y}^{ma}-\mathbf{Y}^{in}, 
\label{eq_Yrel}
\end{equation}
where for the reversible part it directly follows via the corresponding subtraction in Eqs.~(\ref{eq_Yrev}) that 
\begin{equation}
Y^{rel,rev} = a\left[ U^{rel}_{ik}(\nabla_jv_k) + U^{rel}_{jk}(\nabla_iv_k) \right]
  -b^{rel}A_{ij}, 
\end{equation}
with
\begin{equation}
b^{rel} = b^{ma}-b^{in}. 
\end{equation}
Concerning the corresponding reversible parts of the stress components ${\sigma}^{\prime\:rev}_{ij}$ in Eq.~(\ref{eq_sigma-prime-rev}), it is straightforward to show that 
\begin{eqnarray}
\lefteqn{ a\Big[ U^{ma}_{ik}\Phi^{ma}_{kj} + U^{ma}_{jk}\Phi^{ma}_{ki}  
  + U^{in}_{ik}\Phi^{in}_{kj} + U^{in}_{jk}\Phi^{in}_{ki} \Big]  }
\nonumber\\
  &&{}
  - b^{ma}\Phi^{ma}_{ij} - b^{in}\Phi^{in}_{ij}
\nonumber\\
  &=&{} a\Big[ U^{ma}_{ik}\tilde{\Phi}^{ma}_{kj} + U^{ma}_{jk}\tilde{\Phi}^{ma}_{ki}  
  + U^{rel}_{ik}\tilde{\Phi}^{rel}_{kj} + U^{rel}_{jk}\tilde{\Phi}^{rel}_{ki} \Big]  
\nonumber\\
  &&{}
  - b^{ma}\tilde{\Phi}^{ma}_{ij} - b^{rel}\tilde{\Phi}^{rel}_{ij}.
\end{eqnarray}
Therefore, the stress tensor is consistently insensitive to whether we use as macroscopic variables $\mathbf{U}^{ma}$ and $\mathbf{U}^{in}$ with the conjugate variables $\mathbf{\Phi}^{ma}$ and $\mathbf{\Phi}^{in}$, or rather $\mathbf{U}^{ma}$ and $\mathbf{U}^{rel}$ with the conjugate variables $\mathbf{\tilde{\Phi}}^{ma}$ and $\mathbf{\tilde{\Phi}}^{rel}$. 

Finally, we must check whether also the dissipative parts of the quasi-currents comply with Eqs.~(\ref{eq_dynUma})--(\ref{eq_Yrel}). For this purpose, similarly to reconsidering the strain-dependent part of the energy density as in Eq.~(\ref{eq_estr}), we rewrite the $\mathbf{\Phi}$-dependent part of the entropy production rate $R$ in Eq.~(\ref{eq:2R}) as
\begin{equation}
2\tilde{R}^{\Phi} =
  \tilde{\zeta}^{ma}_{ijkl}\tilde{\Phi}^{ma}_{ij}\tilde{\Phi}^{ma}_{kl}
  +2\tilde{\zeta}^{ma,rel}_{ijkl}\tilde{\Phi}^{ma}_{ij}\tilde{\Phi}^{rel}_{kl}
  +\tilde{\zeta}^{rel}_{ijkl}\tilde{\Phi}^{rel}_{ij}\tilde{\Phi}^{rel}_{kl}.
\end{equation}
From this expression, we indeed find
\begin{eqnarray}
Y^{ma,dis}_{ij} &=& \frac{\partial \tilde{R}^{\Phi}}{\partial \tilde{\Phi}_{ij}^{ma}} 
  \nonumber\\[.1cm]
  &=& \tilde{\zeta}^{ma}_{ijkl}\tilde{\Phi}_{kl}^{ma} 
      + \tilde{\zeta}^{ma,rel}_{ijkl}\tilde{\Phi}_{kl}^{rel}
  \nonumber\\[.1cm]
  &=& {\zeta}^{ma}_{ijkl}{\Phi}_{kl}^{ma} 
      + {\zeta}^{ma,in}_{ijkl}{\Phi}_{kl}^{in}
  \nonumber\\[.1cm]
  &=& Y^{ma,dis}_{ij}, 
\\[.1cm]
Y^{rel,dis}_{ij} &=& \frac{\partial \tilde{R}^{\Phi}}{\partial \tilde{\Phi}_{ij}^{rel}} 
  \nonumber\\[.1cm]
  &=& \tilde{\zeta}^{ma,rel}_{ijkl}\tilde{\Phi}_{kl}^{ma} 
      + \tilde{\zeta}^{rel}_{ijkl}\tilde{\Phi}_{kl}^{rel}
  \nonumber\\[.1cm]
  &=& {\zeta}^{ma}_{ijkl}{\Phi}_{kl}^{ma} 
      + {\zeta}^{ma,in}_{ijkl}{\Phi}_{kl}^{in} 
  \nonumber\\[.1cm]
  &&{}
      - {\zeta}^{ma,in}_{ijkl}{\Phi}_{kl}^{ma}
      - {\zeta}^{in}_{ijkl}{\Phi}_{kl}^{in}
  \nonumber\\[.1cm]
  &=& Y^{ma,dis}_{ij} - Y^{in,dis}_{ij}, 
\end{eqnarray}
in agreement with Eq.~(\ref{eq_Yrel}), where we compared with Eqs.~(\ref{eq_Ydis}) in both cases. The following relations between the coefficient tensors result from these comparisons: 
\begin{eqnarray}
\tilde{\zeta}_{ijkl}^{ma} &=& {\zeta}_{ijkl}^{ma}, \\
\tilde{\zeta}_{ijkl}^{ma,rel} &=& {\zeta}_{ijkl}^{ma}-{\zeta}_{ijkl}^{ma,in}, \\
\tilde{\zeta}_{ijkl}^{rel} &=& {\zeta}_{ijkl}^{ma}-2{\zeta}_{ijkl}^{ma,in}
                               +{\zeta}_{ijkl}^{in}. 
\end{eqnarray}

In summary, our analysis in this section demonstrates that \textit{relative strains} can be introduced and used as appropriate macroscopic variables. 
Instead of building our symmetry-based description on solely the absolute strains, involving relative strains is an equally valid approach. That is, working with the macroscopic variables $\mathbf{U}^{ma}$ and $\mathbf{U}^{rel}$ instead of $\mathbf{U}^{ma}$ and $\mathbf{U}^{in}$ throughout our derivation in Secs.~\ref{sec:thermodynamics}--\ref{sec:dynamics} leads to an equivalent formulation of the theory. \revision{Depending on the situation, it can be more intuitive to describe effects that directly originate from strain differences between individual components using the variables of relative strains.} In this way, we have amended the sequence of \textit{relative} macroscopic variables, i.e.\ relative translations and relative rotations, by yet another member in the form of relative strains, \revision{with differences as mentioned below Eq.~(\ref{eq_D2})}.

\section{Minimal examples} \label{sec:examples}

To illustrate 
the background of our description, we consider in the following two minimal example geometries. For this purpose, we concentrate on the central part of our theory, which is the coupling between different strain fields and possible relative strains between them. 

For simplicity, we confine ourselves to linear terms. We neglect changes in density, entropy, and concentration, i.e., we set $\delta\rho=\delta\sigma=\delta c=0$. Moreover, we consider all components of the material to conserve their volume. Thus $U^x_{ii}=0$ ($x\in\{ma,co,in\}$) and $\nabla\cdot\mathbf{v}=0$. 
Under all these assumptions, we neglect relative translations $\mathbf{u}^{rel}$ and relative rotations $\mathbf{\Omega}^{rel}$, and we focus on the strain variables $\mathbf{U}^x$. 



\subsection{Uniaxial extension for elastic matrix, viscoelastic coupling zones, and rigid inclusions}
\label{sec_uniaxial}

As a first example we address the situation outlined already above and observed in recent experiments \cite{gundermann2014investigation}. An elastic permanently crosslinked polymer matrix contains embedded rigid colloidal particles. Due to their rigidity, we do not take into account the strains $\mathbf{U}^{in}$ of these inclusions as a slowly relaxing macroscopic variable. That is, we confine ourselves to the strains $\mathbf{U}^{ma}$ and $\mathbf{U}^{co}$. In the situation of Ref.~\cite{gundermann2014investigation}, it was observed experimentally that the strain deformations of the coupling zones surrounding the inclusions were not completely reversible. 

\begin{figure}
\centerline{\includegraphics[width=6.5cm]{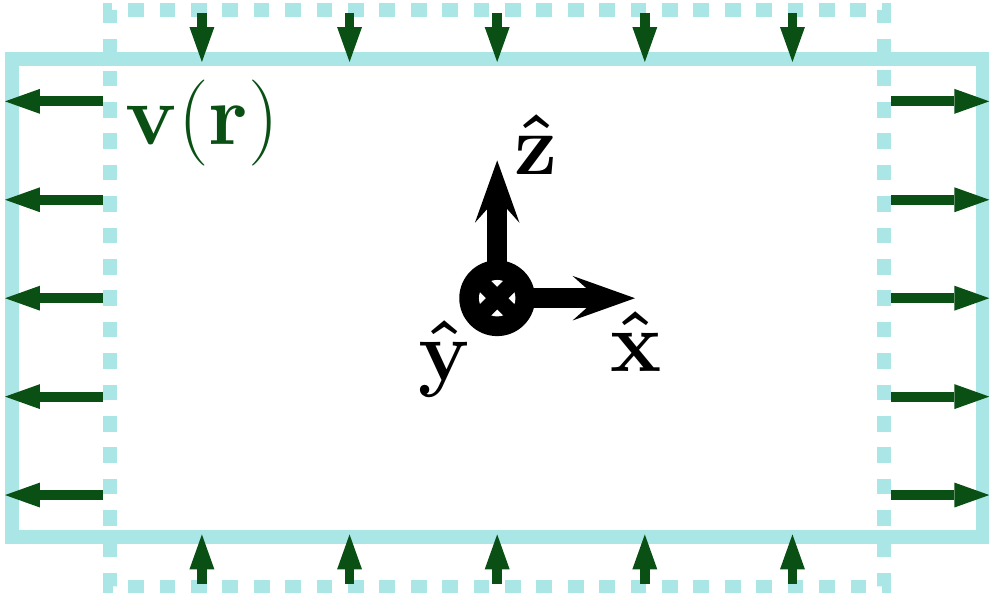}}
\caption{Illustration of our first minimal example geometry: a spatially homogeneous uniaxial extension along the $x$ direction. The deformation is driven by a volume-conserving flow field $\mathbf{v}(\mathbf{r})$ as indicated by the arrows and given by Eq.~(\ref{eq_At}). To preserve its volume, the system must contract from the sides, i.e.\ along the $y$ and $z$ directions. 
}
\label{fig_magneticgels}
\end{figure}

For simplicity, we consider spatially homogeneous uniaxial extensions along the $x$ direction of our Cartesian coordinate system, see Fig.~\ref{fig_magneticgels},
imposed by a corresponding velocity field
of the type 
\begin{equation}
\mathbf{v}(\mathbf{r}) = \left( \begin{array}{c} A(t)x \\ -\frac{1}{2}A(t)y \\[.1cm] -\frac{1}{2}A(t)z \end{array} \right).  
\label{eq_At}
\end{equation}
$A(t)$ sets the time-dependent amplitude of the flow field. To preserve its volume, the system must contract from the sides, leading to the contributions along the $y$ and $z$ directions. 

We assume both matrix and coupling zones to deform instantaneously with our imposed flow fields, i.e.\ $b^{ma}\neq0\neq b^{co}$ in Eqs.~(\ref{eq_Yrev}). In both cases, this leads to stresses $\mathbf{\Phi}^x=c_1^x\mathbf{U}^x$, see Eqs.~(\ref{eq_cx}) and (\ref{eq_Phi}), where we only consider elastic coefficients $c_1^x\neq0$ ($x\in\{ma,co\}$). The matrix can only react reversibly and elastically to the externally imposed deformations, $\zeta_1^{ma}=0$, whereas the stresses within the less crosslinked coupling zones can decay through irreversible rearrangements, e.g.\ disentanglements, thus $\zeta_1^{co}\neq0$ in Eqs.~(\ref{eq_zetax}) and (\ref{eq_Ydis}). 

Under all these assumptions, the remaining dynamic equations under consideration resulting from Eqs.~(\ref{eq_dynU}) read
\begin{eqnarray}
\partial_t\mathbf{U}^{ma} &=& b^{ma}\mathbf{A}, \\
\partial_t\mathbf{U}^{co} &=& b^{co}\mathbf{A} - c_1^{co}\zeta_1^{co}\mathbf{U}^{co},
\end{eqnarray}
where
\begin{equation}
\mathbf{A} = \left( \begin{array}{ccc} 
             A(t) & 0 & 0 \\
             0 & -\frac{1}{2}A(t) & 0 \\
             0 & 0 & -\frac{1}{2}A(t) \end{array} \right).
\end{equation}
The stress $\bm{\sigma'}$ resulting via Eq.~(\ref{eq_sigma-prime-rev}) from the described deformations reads
\begin{equation}
\bm{\sigma'} = {}-b^{ma}\mathbf{\Phi}^{ma} -b^{co}\mathbf{\Phi}^{co}
               -\nu_1\mathbf{A}. 
\end{equation}
This is the stress exerted by the system. In other words, the applied stress necessary to achieve the deformations is $-\bm{\sigma'}$. 

For simplicity, we here set $c_1^{co}=c_1^{ma}$ as well as $b^{co}=b^{ma}$, and we measure all times in units of $1/c_1^{ma}\zeta_1^{ma}$, the velocity amplitude $A(t)$ in units of $c_1^{ma}\zeta_1^{ma}/b^{ma}$, the stress $\bm{\sigma'}$ in units of $c_1^{ma}b^{ma}$, as well as the viscosity $\nu_1$ in units of $(b^{ma})^2/\zeta_1^{ma}$.

\begin{figure}
\centerline{\includegraphics[width=8.5cm]{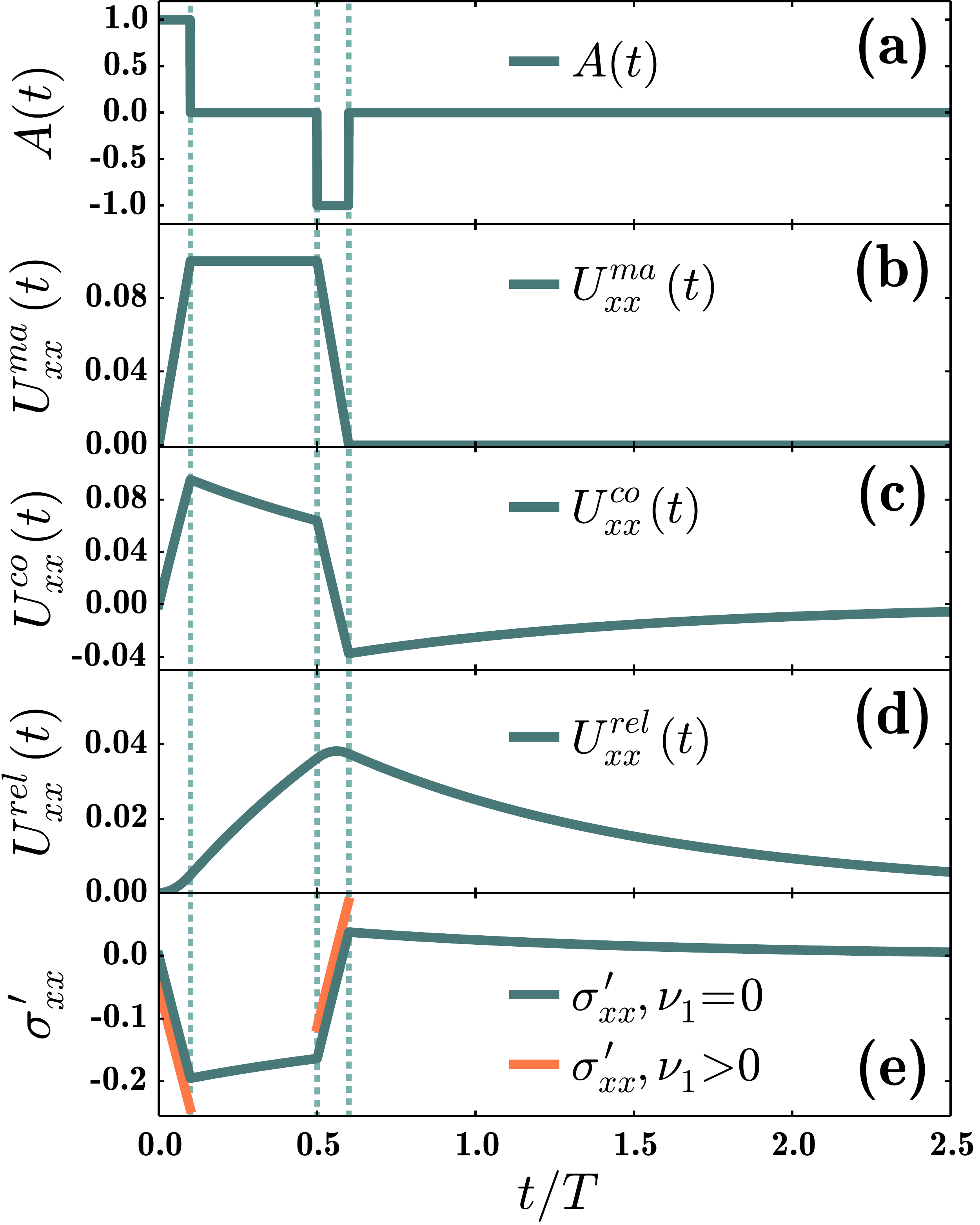}}
\caption{Results for the  uniaxial extension illustrated in Fig.~\ref{fig_magneticgels}. A velocity field $\mathbf{v}(\mathbf{r})$ [see Fig.~\ref{fig_magneticgels} and Eq.~(\ref{eq_At})] is applied with an amplitude $A(t)$ as displayed in panel (a) showing two opposite rectangular pulses. (b) The matrix deformation is completely reversible, see Eq.~(\ref{eq_magneticgels_Uma}). (c) In contrast to that, exponential relaxation of the strain takes place in the coupling zones, see Eq.~(\ref{eq_magneticgels_Uco}), leading to the inverted strain upon the impact of the sudden inverted pulse of deformation. (d) From the different behaviors of the two components, relative strains $\mathbf{U}^{rel}=\mathbf{U}^{ma}-\mathbf{U}^{co}$ arise between them. (e) Likewise, the signs of the stresses change upon the inverse second pulse, leading to oppositely oriented external forces necessary to keep the system in its prescribed state. If the viscosity $\nu_1$ is non-zero (here $\nu_1=0.05$), additional stresses result for non-vanishing velocities $\mathbf{v}(\mathbf{r})$. The vertical dashed lines mark the times $t_p$ after which the first pulse ends, $T/2$ when the second pulse is switched on, and $T/2+t_p$ after which the second pulse ends. 
}
\label{fig_magneticgels_results}
\end{figure}

We start our considerations at time $t=0$ and impose two rectangular pulses onto the velocity amplitude, 
\begin{equation}
A(t) = \left\{
\begin{array}{cl}
A, & 0\leq t<t_p, \\[.1cm]
0, & t_p\leq t<\frac{1}{2}T, \\[.1cm]
-A, & \frac{1}{2}T\leq t<\frac{1}{2}T+t_p, \\[.1cm]
0, & \frac{1}{2}T+t_p\leq t,
\end{array}
\right.
\label{eq_Aprescr}
\end{equation} 
with $A$ the constant amplitude, $t_p$ the duration of each pulse, and $T/2$ the time between the onsets of the two pulses, see also Fig.~\ref{fig_magneticgels_results}~(a).

For the time intervals given in Eq.~(\ref{eq_Aprescr}) and using our rescaled units, we obtain, respectively, 
\begin{equation}
U^{ma}_{xx} = \left\{
\begin{array}{c}
At, \\[.1cm]
At_p, \\[.1cm]
A\left(\frac{1}{2}T+t_p-t\right), \\[.1cm]
0,
\end{array}
\right.
\label{eq_magneticgels_Uma}
\end{equation}
together with
\begin{equation}
U^{co}_{xx} = \left\{
\begin{array}{l}
A\big( 1-e^{-t} \big), \\[.1cm]
A\big( e^{-(t-t_p)}-e^{-t} \big), \\[.1cm]
A\big( e^{-(t-t_p)}-e^{-t}  
  -1 + e^{-(t-\frac{1}{2}T)} \big), \\[.1cm]
A\big( e^{-(t-t_p)}-e^{-t} \\[.1cm]
  \qquad {}-e^{-(t-(\frac{1}{2}T+t_p))} 
  + e^{-(t-\frac{1}{2}T)} \big).
\end{array}
\right.
\label{eq_magneticgels_Uco}
\end{equation}
These quantities, together with the resulting relative strain $U_{xx}^{rel}=U_{xx}^{ma}-U_{xx}^{co}$ and the stress components $\sigma'_{xx}$, are plotted in Fig.~\ref{fig_magneticgels_results}~(b)--(e). Corresponding expressions for the $y$ and $z$ directions follow with $A$ replaced by $-A/2$. 

The results depicted in Fig.~\ref{fig_magneticgels_results}~(b)--(e) allow a very illustrative interpretation. When the material is deformed due to the imposed velocity field, both matrix and coupling zones directly and instantaneously deform accordingly. The matrix reacts in a completely elastic and reversible way. That is, its strain deformation can only be reversed by imposing the inverse velocity field during the second, inverse pulse [Fig.~\ref{fig_magneticgels_results}~(b)]. In contrast to that, the strain of the coupling zones decays via irreversible processes [Fig.~\ref{fig_magneticgels_results}~(c)], e.g.\ via disentanglement of polymer chains. This leads to relative strains between the two components [Fig.~\ref{fig_magneticgels_results}~(d)] and reduces the necessary stress to keep the whole material in the overall strained state [Fig.~\ref{fig_magneticgels_results}~(e)]. When the strain of the matrix is reversed during the second, inverse pulse, the coupling zones have to some extent already relaxed their initial strain deformation. 
Then, the inverse pulse that reverses and releases the matrix strain leads to an effective compression of the coupling zones [Fig.~\ref{fig_magneticgels_results}~(c)]. We remark that the relative strain between the two components remains approximately constant during this short pulse [Fig.~\ref{fig_magneticgels_results}~(d)]. The compressed coupling zones now oppose to the strain-released state of the matrix. Thus the overall stress switches its sign [Fig.~\ref{fig_magneticgels_results}~(e)]. A compressive stress is necessary to keep the overall material in the state of released matrix strain. Afterwards, this stress again decays due to the decay of strain in the coupling zones [Fig.~\ref{fig_magneticgels_results}~(c)]. 

\revision{As demonstrated in Sec.~\ref{sec:relstrains}, approaching the situation either via $\mathbf{U}^{ma}$ and $\mathbf{U}^{co}$ [Fig.~\ref{fig_magneticgels_results}~(b) and (c)] or via $\mathbf{U}^{ma}$ and $\mathbf{U}^{rel}$ [Fig.~\ref{fig_magneticgels_results}~(b) and (d)] is completely equivalent. In some respects, the description using relative strains is more illustrative. The relative strains simply decay too slowly to change significantly while the inverse pulse is applied [Fig.~\ref{fig_magneticgels_results}~(d)]. This is why the less stretched coupling zones are driven into the compressed state. Afterwards, the relative strains slowly relax.}



\subsection{Poiseuille flow for (visco)elastic composite materials}

\begin{figure}
\centerline{\includegraphics[width=6.5cm]{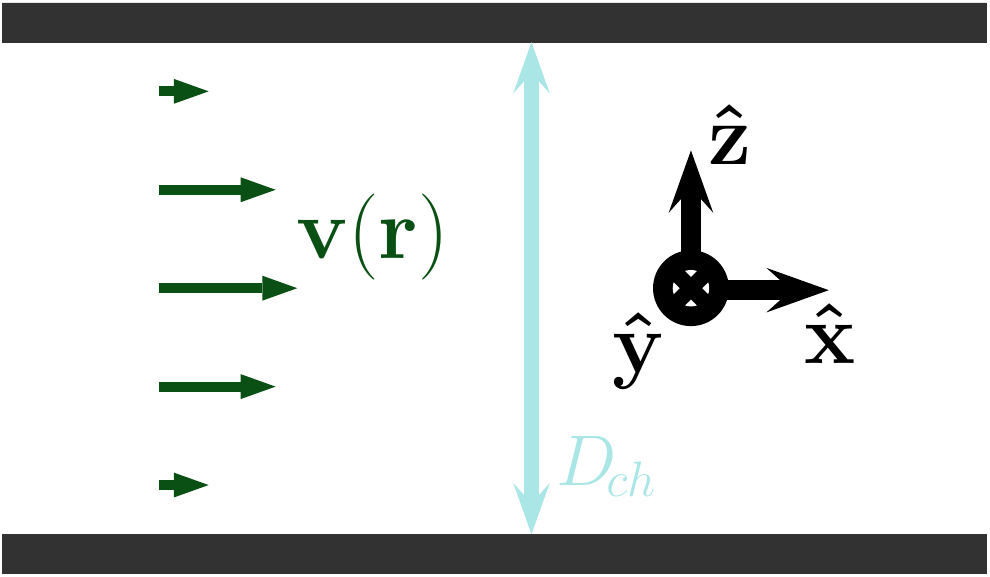}}
\caption{Illustration of our second minimal example geometry: a Poiseuille channel flow. Two parallel plates are separated by a distance $D_{ch}$. The inner surfaces of these plates are located at $z=\pm D_{ch}/2$. On these plate surfaces, we impose $\mathbf{v}=\mathbf{0}$. In contrast to that, free slippage is allowed on confining plates with normals along the $y$ direction. A flow $\mathbf{v}(\mathbf{r})$ is excited, possibly by an externally applied pressure gradient parallel to the $x$ direction. 
}
\label{fig_poiseuille_geometry}
\end{figure}

As a second minimal example, we consider the geometry of a Poiseuille channel geometry, see Fig.~\ref{fig_poiseuille_geometry}. 
There, the material is confined between two parallel plates of separation distance $D_{ch}$. 
The plates are oriented with their normals along the $z$ direction of our Cartesian coordinate system. Vanishing velocity ($\mathbf{v}=\mathbf{0}$) is imposed on the plate surfaces, that is for $z=\pm D_{ch}/2$. In contrast to that, we assume the system to be effectively confined in the $y$ direction by rigid walls that allow slippage ($\mathbf{v}\neq\mathbf{0}$) on their surfaces without any friction. 
Along the $x$ direction, which will be our flow direction, the system is of significantly larger dimension than $D_{ch}$. Boundary effects at the channel ends will not be taken into account. 

Overall, with all these assumptions, the flow of the material is considered to be confined to the $x$ direction, spatially homogeneous in $x$ and $y$ directions. That is, $\mathbf{v}(\mathbf{r},t)=v_x(z,t)\mathbf{\hat{x}}$. 
Next, we apply from outside a time-dependent pressure gradient $(\nabla p)(t)=(\partial_xp)(t)\mathbf{\hat{x}}$ along the channel parallel to the $x$ direction. Under our assumptions, this pressure gradient shall be spatially homogeneous within our material. Alternatively, one may think of a spatially homogeneous gravitational bulk volume force. Its influence on the material can be switched by turning the system in the gravitational field. As a consequence, the only components of the strain tensors to be taken into account are the $U^r_{xz}=U^r_{zx}$ shear components ($r\in\{ma,co,in\}$; to avoid confusion with the spatial coordinates we here switch to superscripts $r$ and $s$). 

Using our assumptions and simplifications, we infer the dynamic equations for the velocity $v_x$ and shear strains $U^r_{zx}$ from Eqs.~(\ref{eq_cx}), (\ref{eq_cxy}), (\ref{eq_Phi}), (\ref{eq_dyng}), (\ref{eq_dynU}), (\ref{eq_sigma-sigmaprime-p}), (\ref{eq_sigmasplit}), (\ref{eq_Ysplit}), (\ref{eq_Yrev}), (\ref{eq_sigma-prime-rev}), (\ref{eq_nu})--(\ref{eq_zetaxy}), (\ref{eq_sigmadis}), and (\ref{eq_Ydis}). 
Let us first briefly address the case of $\zeta_1^r=\zeta_1^{r,s}=0$. That is, there are no irreversible contributions to the strain deformations via the dissipative quasi-currents in Eqs.~(\ref{eq_Ydis}). 

Measuring time in units of $\nu_1/c_1^{ma}$, space in units of $\nu_1/\sqrt{\rho c_1^{ma}}$, velocity in units of $\sqrt{c_1^{ma}}/\sqrt{\rho}b^{ma}$, 
pressure in units of $c_1^{ma}/b^{ma}$, as well as $c_1^{r(,s)}$ and $b^r$ in units of $c_1^{ma}/(b^{ma})^2$ and $b^{ma}$, respectively, 
and as a consequence stresses $\mathbf{\Phi}^r$ in units of $c_1^{ma}/(b^{ma})^2$, 
we find the dynamic equations
\begin{eqnarray}
\partial_tv_x &=& {}-\partial_xp + \sum_rb^r\partial_z\Phi_{zx}^r + \frac{1}{2}\partial_z^2v_x, 
\label{poiseuille_velastic}
\\
\partial_tU^r_{zx} &=& \frac{1}{2}b^r\partial_zv_x, 
\label{poiseuille_Uelastic}
\end{eqnarray}
where 
\begin{equation}
\Phi_{zx}^r = c_1^rU^r_{zx} + \sum_{s\neq r}c_1^{r,s}U_{zx}^s,
\label{poiseuille_Phielastic}
\end{equation}
$b^{ma}=c_1^{ma}=1$, and $r,s\in\{ma,co,in\}$. 

\begin{figure}
\centerline{\includegraphics[width=8.5cm]{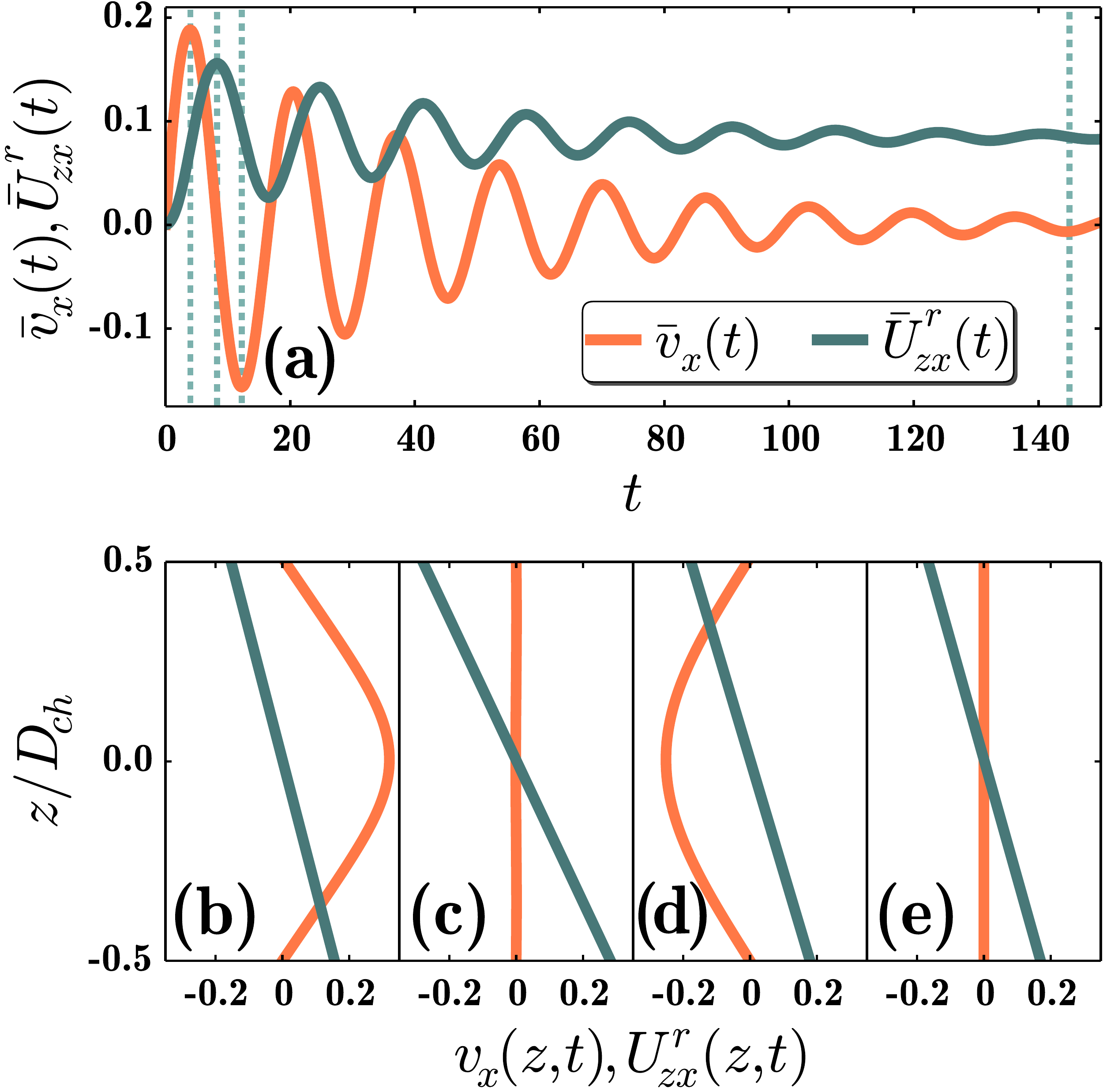}}
\caption{Response of the system illustrated in Fig.~\ref{fig_poiseuille_geometry} to an external pressure gradient $\partial_xp=-0.1$ suddenly switched on at $t=0$. The strain deformations are completely reversible in this case, i.e.\ $\zeta_1^r=\zeta_1^{r,s}=0$, such that the system is described by Eqs.~(\ref{poiseuille_velastic})--(\ref{poiseuille_Phielastic}). For simplicity, we set $c_1^r=1=b^r$ and $c_1^{r,s}=0$ ($r,s\in\{ma,co,in\}$). (a) Time evolution of the quantities $\bar{v}_x(t)$ and $\bar{U}^r_{zx}(t)$ averaged across the channel as described in the text. The spatial profiles for $v_x(z,t)$ and $U^r_{zx}(z,t)$ are plotted at different times (b) $t=4$, (c) $t=8.265$, (d) $t=12.25$, and (e) $t=145$, as indicated by the dotted lines in panel (a). 
}
\label{fig_poiseuille_elastic}
\end{figure}

Fig.~\ref{fig_poiseuille_elastic} illustrates the process of suddenly switching on an external pressure gradient $\partial_xp=-0.1$ at time $t=0$ in a system previously at rest. 
For simplicity, we here set in this qualitative demonstration $c_1^r=1=b^r$ and $c_1^{r,s}=0$ ($r,s\in\{ma,co,in\}$), which is why the strain curves for different $r$ collapse. In Fig.~\ref{fig_poiseuille_elastic}~(a), we plot the time evolution of the velocity component $v_x(z,t)$ averaged over the channel width, i.e.\ $\bar{v}_x(t)=\int_{-D_{ch}/2}^{D_{ch}/2}v_x(z,t)\,\mathrm{d}z/D_{ch}$. Since the strain deformations $U^r_{zx}(z,t)$ are antisymmetric with respect to the channel center at $z=0$, we instead use $\bar{U}^r_{zx}(t)=\int_{-D_{ch}/2}^{0}\left[U^r_{zx}(z,t)-U^r_{zx}(-z,t)\right]\mathrm{d}z/D_{ch}$ to quantify the time evolution of the overall strains in the system. 

The negative pressure gradient tends to push the system to the right in the set-up depicted in Fig.~\ref{fig_poiseuille_geometry}. Its sudden application at $t=0$ leads to an acceleration of the system, inducing the initial growth of $\bar{v}_x(t)$ in Fig.~\ref{fig_poiseuille_elastic}~(a). A snapshot of the velocity and strain profiles is depicted in Fig.~\ref{fig_poiseuille_elastic}~(b). However, the growing strains in the system due to the no-slip boundary conditions at $z=\pm D_{ch}/2$ lead to counteracting decelerating stresses via Eqs.~(\ref{poiseuille_Phielastic}). As a consequence, the system comes to rest at maximum overall distortion, see Fig.~\ref{fig_poiseuille_elastic}~(a) and the corresponding profiles in Fig.~\ref{fig_poiseuille_elastic}~(c). Obviously, the system overshoots a possible balance between the external pressure gradient and the induced stresses as found at later times. The velocity then reverses, see Fig.~\ref{fig_poiseuille_elastic}~(a) and associated profiles in Fig.~\ref{fig_poiseuille_elastic}~(d). In this way, the system oscillates around the new equilibrium point. Yet, the oscillation amplitude decreases due to the viscous damping described by the last term in Eq.~(\ref{poiseuille_velastic}). Finally, the system approaches a static distorted state close to the profile depicted in Fig.~\ref{fig_poiseuille_elastic}~(e).
In analogy, when the external pressure gradient is suddenly switched off again at a later time, all strains reverse to zero in an oscillatory fashion (not shown). 

Next, we allow for dissipative processes described by the coefficients $\zeta_1^{r(,s)}$ in Eqs.~(\ref{eq_zetax}), (\ref{eq_zetaxy}), and (\ref{eq_Ydis}). That is, induced strains may decay due to irreversible processes. An example process is disentangling of polymer chains \cite{degennes1979scaling,doi2007theory} in polymer melts or weakly crosslinked polymeric systems. 

Measuring time in units of $1/c_1^{ma}\zeta_1^{ma}$, space in units of $b^{ma}/\zeta_1^{ma}\sqrt{\rho c_1^{ma}}$, velocity in units of $\sqrt{c_1^{ma}/\rho}$, 
pressure in units of $c_1^{ma}b^{ma}$, the viscosity $\nu_1$ in units of $(b^{ma})^2/\zeta_1^{ma}$, as well as $c_1^{r(,s)}$, $\zeta_1^{r(,s)}$, and $b^r$ in units of $c_1^{ma}$, $\zeta_1^{ma}$, and $b^{ma}$, respectively, and as a consequence stresses $\mathbf{\Phi}^r$ in units of $c_1^{ma}$,  
we find the dynamic equations
\begin{eqnarray}
\partial_tv_x &=& {}-\partial_xp + \sum_rb^r\partial_z\Phi_{zx}^r + \frac{1}{2}\nu_1\partial_z^2v_x, 
\label{poiseuille_v}
\\
\partial_tU^r_{zx} &=& \frac{1}{2}b^r\partial_zv_x -\zeta_1^r\Phi_{zx}^r -\sum_{s\neq r}\zeta_1^{r,s}\Phi_{zx}^s, 
\label{poiseuille_U}
\end{eqnarray}
where 
\begin{equation}
\Phi_{zx}^r = c_1^rU^r_{zx} + \sum_{s\neq r}c_1^{r,s}U_{zx}^s,
\label{poiseuille_Phi}
\end{equation}
$b^{ma}=\zeta_1^{ma}=c_1^{ma}=1$, and $r,s\in\{ma,co,in\}$. 

At this point, let us turn to an example that shows the limitations of the present formulation of our approach. Yet, this simple example also reveals how the present formulation must be extended to include still more general situations, and which practical cases are represented by our current description. 

\begin{figure}
\centerline{\includegraphics[width=8.5cm]{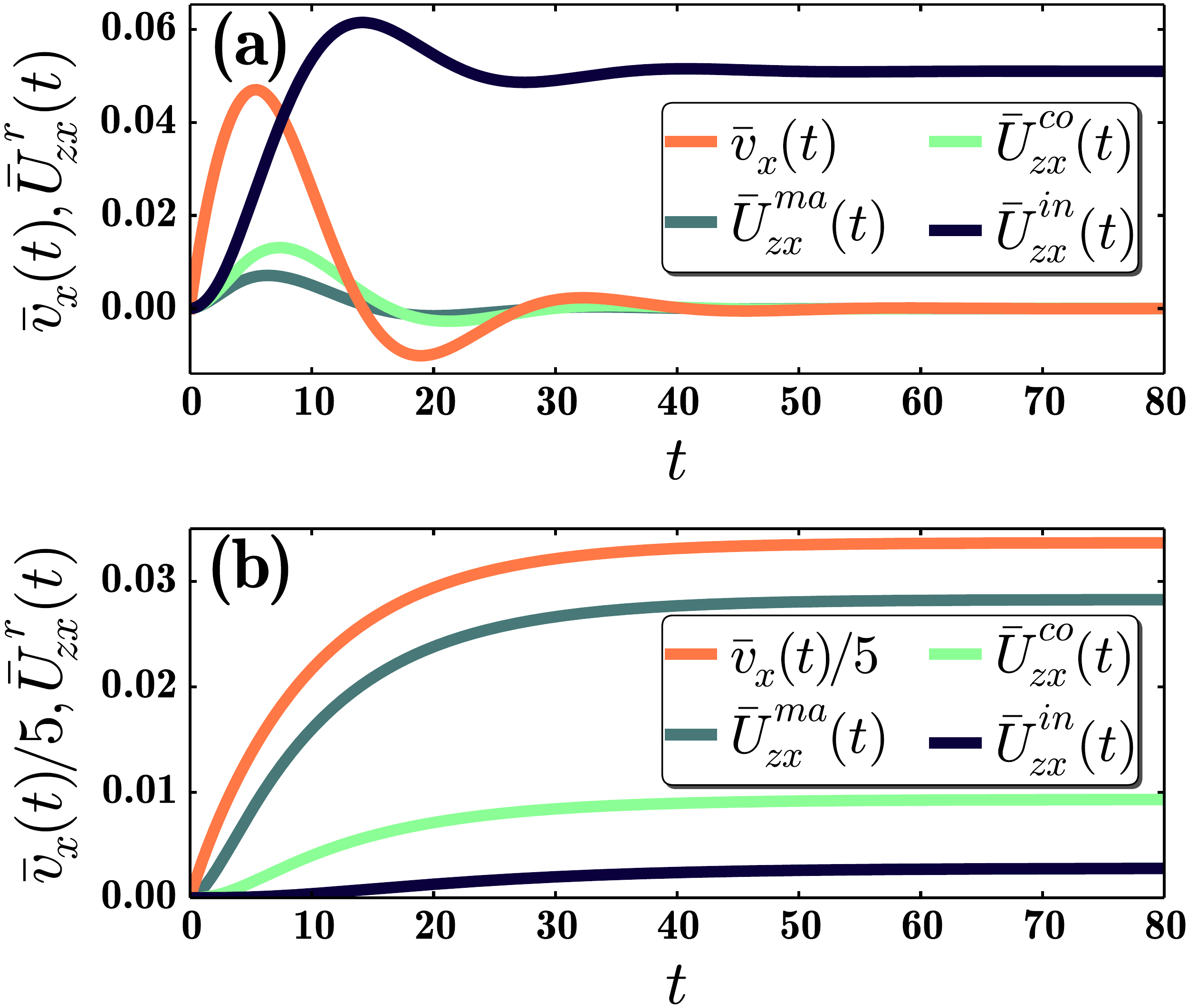}}
\caption{Response of the set-up in Fig.~\ref{fig_poiseuille_geometry} to an external pressure gradient $\partial_xp=-0.02$ suddenly switched on at $t=0$. The system is described by Eqs.~(\ref{poiseuille_v})--(\ref{poiseuille_Phi}), where we set $\zeta_1^{r,s}=0$ and $\nu_1=1$ ($r,s\in\{ma,co,in\}$). In the first case (a) the flow field is directly coupled to all zones ($b^r=1$), while the ``inclusions'' respond in a perfectly elastic reversible way ($\zeta_1^{ma}=1$, $\zeta_1^{co}=0.5$, $\zeta_1^{in}=0$, $c_1^r=1$, $c_1^{r,s}=0$). Then the overall flow finally ceases. In the second case (b) there is no direct coupling between the flow field and the coupling zones or inclusions ($b^{ma}=1$, $b^{co}=b^{in}=0$). The non-vanishing coefficient $\zeta_1^{in}=0.1$ here is interpreted to result from an overdamped deformation process of the inclusions due to their coupling to their environment ($\zeta_1^{ma}=1$, $\zeta_1^{co}=0.5$). Strains are energetically transmitted from the matrix via the coupling zones to the inclusions via $c_1^{ma,co}=c_1^{co,in}=-0.3$ ($c_1^{ma,in}=0$, $c_1^r=1$). A terminal flow develops in this situation, with non-vanishing strains of all zones. 
}
\label{fig_poiseuille_fail}
\end{figure}

We again consider a scenario where at $t=0$ the external pressure gradient is suddenly switched on to a non-vanishing value, here $\partial_xp=-0.02$. 
Once more, we set for simplicity $c_1^r=1=b^r$ and $c_1^{r,s}=\zeta_1^{r,s}=0$ ($r,s\in\{ma,in,co\}$). However, the coefficients controlling the dissipative quasi-currents in the dynamic strain equations are now chosen as $\zeta^{ma}=1$, $\zeta^{co}=0.5$, and $\zeta^{in}=0$. The time evolution of the resulting velocity and strains averaged across the channel are shown in Fig.~\ref{fig_poiseuille_fail}~(a). 

In view of the chosen values for $\zeta_1^r$, we might think, for instance, of a non-permanently crosslinked polymer matrix that contains localized elastic inclusions. In that case, our intuition would guide us to an ultimately non-vanishing net flow through the channel, down the external pressure gradient. This flow should be caused by the continuous disentangling processes within the polymer matrix described by $\zeta_1^{ma}\neq0$. However, we observe in Fig.~\ref{fig_poiseuille_fail}~(a) that the flow ceases after a transient time, i.e.\ $v_x(z,t\rightarrow\infty)=0$. In our equations, we find the underlying reason in the built-up of stresses in the inclusions as a consequence of the non-decaying strains $U^{in}_{zx}$ ($\zeta_1^{in}=0$), see Fig.~\ref{fig_poiseuille_fail}~(a) and Eqs.~(\ref{poiseuille_v})--(\ref{poiseuille_Phi}). The more the system would flow, the more these stresses build up and counteract the flow, just as in our previous case for the wholly elastic system in Fig.~\ref{fig_poiseuille_elastic}. In contrast to what one would expect in reality, the built-up of stress $\mathbf{\Phi}^{in}$ here ceases the flow. 

Why does our description fail in the presented situation? The reason is that we formulated our approach only for strongly coupled situations that can be described by one overall momentum density $\mathbf{g}$. Instead, we would now need for an additional macroscopic symmetrized tensor variable $\mathbf{A}^{in}$ of second rank that can drive and reverse elastic distortions of the inclusions, also relatively to the overall flow. 
It is the analogue to our globally defined $\mathbf{A}$ in the present formalism, the components of which were given by Eq.~(\ref{eq_Aij}). Further remarks on this point are included in Sec.~\ref{sec:discussion}, and a corresponding theory shall be developed in the future. 
In contrast to that, an example system, where the features described by Fig.~\ref{fig_poiseuille_fail}~(a) should be relevant, are entangled (semi-)interpenetrating polymer networks \cite{sperling1996current,myung2008progress}. 
Or, for clearly separated time scales, we simply neglect $\mathbf{U}^{in}$ as a variable,  
similarly to our procedure for rigid inclusions in Sec.~\ref{sec_uniaxial}, which then allows for a steady net flow. 

\begin{figure*}
\centerline{\includegraphics[width=17.8cm]{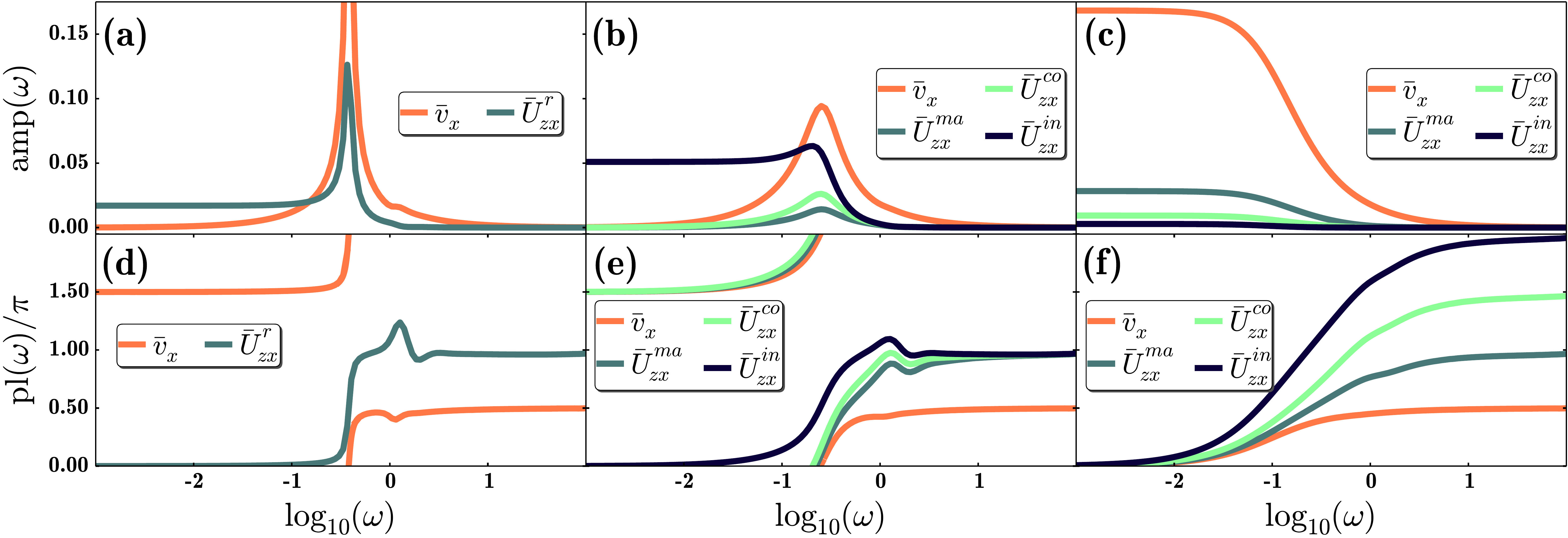}}
\caption{Rheological response of the systems introduced in Figs.~\ref{fig_poiseuille_elastic}, \ref{fig_poiseuille_fail}~(a), and \ref{fig_poiseuille_fail}~(b) to an oscillating external pressure gradient $(\partial_xp)(t)=0.02\sin(\omega t)$. The first system is characterized by completely reversible strain deformations, the second one features dissipative contributions to the strain quasi-currents via the coefficients $\zeta_1^r$, while in the third system the components $co$ and $in$ are indirectly connected to the external stimulus only by energetic couplings via the dissipative quasi-currents set by $\zeta_1^r$ ($r\in\{ma,co,in\}$). In this order, the results for the three different systems are depicted from left to right in the different panels for the quantities indicated by the figure legends. As a function of frequency $\omega$, (a)--(c) show the corresponding oscillation amplitudes $\mathrm{amp}(\omega)$, while (d)--(f) give the resulting phase lags $\mathrm{pl}(\omega)$ of the responses, all shifted to the interval $[0,2\pi[$. (We plot the phase lags with respect to $-(\partial_xp)(t)$ because a static $\partial_xp<0$ leads to a flow into positive $x$ direction.)
}
\label{fig_rheology}
\end{figure*}

Nevertheless, from a practical point of view, we should not be too pejorative. For many practical situations, our present description will still be sufficient, if we may represent the situation in the following way. First, we disconnect in our linearized formulation the overall flow field $\mathbf{v}$ from the stresses in the coupling zones and inclusions, i.e.\ we set $b^{co}=b^{in}=0$. Thus the flow field $\mathbf{v}$ is mainly connected to the matrix flow, which for a bulk matrix and a sufficiently low concentration of localized inclusions should represent a reasonable approximation. Next, we do allow for non-vanishing $\zeta_1^{co}$ and $\zeta_1^{in}$, even if the isolated inclusions only deform elastically in a reversible way. In this situation, $\zeta_1^{in}\neq0$ does not imply irreversible topological processes inside the inclusions, such as disentanglements of polymer chains. Rather, it represents the overdamped character of their deformation kinetics, for instance due to the embedding into and coupling to highly viscous coupling zones or matrix environments. Finally, deformations are now transmitted from the matrix via the coupling zones to the inclusions (and vice versa) by (quasi-)static energetic couplings, here given by elastic coefficients $c_1^{ma,co}<0$ and $c_1^{co,in}<0$. Illustratively, this means that the external pressure gradient generates flow and resulting distortions in the matrix. If the matrix is deformed, it is energetically most favorable for the coupling zones to deform in a similar way, at least to a certain degree. If the coupling zones are deformed, these strains are also partially transmitted to the inclusions. Fig.~\ref{fig_poiseuille_fail}~(b) shows the corresponding process, leading to a net steady flow and induced strains of all zones, just as what we would expect in this situation. 

With all these insights at hand, we can further think of rheological applications. In our context of a Poiseuille channel geometry, we may consider a periodically oscillating external pressure gradient $(\partial_xp)(t)$. Alternatively, a periodic tilting within the gravitational field would serve similar purpose.  

As an illustration, we here briefly present results for the three different systems introduced in Figs.~\ref{fig_poiseuille_elastic}, \ref{fig_poiseuille_fail}~(a), and \ref{fig_poiseuille_fail}~(b). They are characterized by the material parameters as given in the corresponding figure captions. An oscillating external pressure gradient of the form $(\partial_xp)(t)=0.02\sin(\omega t)$ is applied. 

For the three different systems, we numerically determined the oscillation amplitudes as well as the phase lags of their oscillations for the four quantities $\bar{v}_x(t)$ and $\bar{U}^r_{zx}(t)$ ($r\in\{ma,co,in\}$). The results are plotted as a function of frequency $\omega$ in Fig.~\ref{fig_rheology}. 
As may have been expected, the system in Fig.~\ref{fig_poiseuille_elastic} characterized by completely reversibly strains and vanishing deformational damping ($\zeta_1^{r(,s)}=0$) shows a pronounced response at a certain resonance frequency, see Fig.~\ref{fig_rheology}~(a). The resonance is reflected by a significant variation in the phase shift of the response around the corresponding frequency, see Fig.~\ref{fig_rheology}~(d). Switching on the damping via the coefficients $\zeta_1^r$ for the two other systems [see Fig.~\ref{fig_rheology}~(b) and (c)] naturally reduces resonance. As likewise may have been expected, the systems (quasi-)statically follow the externally imposed pressure gradient at very low frequencies. They show vanishing response when they cannot follow the external stimulus any more at too high frequencies. Particularly for the third system, 
the increasing phase lag in Fig.~\ref{fig_rheology}~(f) from $\bar{v}_x(t)$ to $\bar{U}^{ma}_{zx}(t)$ to $\bar{U}^{co}_{zx}(t)$ to $\bar{U}^{in}_{zx}(t)$ nicely illustrates how the external stimulus is sequentially handed over from quantity to quantity via their mutual couplings.

\section{Discussion} \label{sec:discussion}

When we developed our approach, we included three zones of possibly different strain deformations. As mentioned in Sec.~\ref{sec:introduction}, this was, for instance, inspired by the observation in magnetic gels of a coupling zone around the inclusions that may have (visco)elastic properties markedly different from the bulk of the matrix material \cite{gundermann2014investigation,huang2016buckling}. Depending on the particular situation, this phenomenological division into three zones may naturally be adjusted. On the one hand, if a still more continuous transition between the properties of the inclusions and the bulk matrix needs to be modeled, the effect of even more intermediate strain variables may be included. On the other hand, as we noted in Secs.~\ref{sec:relstrains} and \ref{sec:examples}, it might not be necessary to include an explicit additional strain variable for all three zones. 
For instance, when rigid metallic particles are embedded in a polymer matrix, it can be conceivable to neglect the strain variable $\mathbf{U}^{in}$ in the (slow) macroscopic dynamics. 

During our presentation, one may have noticed that the couplings between our variables that excite relative translations or rotations are relatively sparse. Nevertheless, we included these variable because they will become important in more complex situations. 
Here, we found that relative translations can be induced by gradients in the temperature or relative chemical potential via the dissipative current in Eq.~(\ref{eq_qdis}). 
Another physical way would be to create 
phase shifts between the displacement dynamics 
of different components of the material. For instance, shaking the whole block of material up and down may induce an out-of-phase swinging of the inclusions against the matrix. In our present hydrodynamic approach, such situations are not included. We only use one variable for the (overall) momentum density $\mathbf{g}$. 
This implies a pronounced coupling between the different components of the material. 
In particular, overdamped and irreversible processes 
favor such restrictions. To resolve more decoupled problems, where individual components may swing relatively to each other, separate momentum densities for the different components need to be introduced. Yet, as becomes clear from the present study, in addition to separate velocity fields $\mathbf{v}^{co}$ and $\mathbf{v}^{in}$ for the coupling zones and inclusions, also separate dynamic fields $\bm{\omega}^{co}$ and $\bm{\omega}^{in}$ as well as $\mathbf{A}^{co}$ and $\mathbf{A}^{in}$ for their rotations and strains, respectively, may be necessary. Similarly to the strains $\mathbf{U}^{co}$ and $\mathbf{U}^{in}$, which in general cannot be obtained from coarse-grained macroscopic displacement fields $\mathbf{u}^{co}$ and $\mathbf{u}^{in}$, see Sec.~\ref{sec:variables} and Fig.~\ref{fig:no_inclusion_displfield}, $\mathbf{A}^{co}$ and $\mathbf{A}^{in}$ in general cannot be obtained from coarse-grained macroscopic velocity fields $\mathbf{v}^{co}$ and $\mathbf{v}^{in}$. Establishing corresponding two- or three-fluid hydrodynamic descriptions \cite{pleiner2004general,pleiner2004generalaip,pleiner2013active, puljiz2016thermophoretically} for (visco)elastic composite materials will be an interesting subject for the future. 

Finally, a different possibility to induce relative rotations arises for anisotropic inclusions. These may be selectively reoriented using externally applied fields. For instance, in liquid crystal elastomers \cite{brand2006selected,urayama2007selected,menzel2015tuned}, relative rotations were generated via external electric fields acting on the liquid crystalline component in a swollen state \cite{yusuf2005low,urayama2006deformation,urayama2007selected,hashimoto2008multifunctional}. Likewise, anisotropic magnetic inclusions could be selectively reoriented by magnetic fields \cite{pessot2015towards,roeder2015magnetic}. We will report on a first application of our formalism to magnetic gels elsewhere.

\section{Conclusions} \label{sec:conclusions}

In this work, we described the behavior of (visco)elastic composite materials using a macroscopic hydrodynamic approach based on symmetry arguments. We integrated into our characterization \textit{relative translations} and \textit{relative rotations} between a matrix and embedded inclusions. Moreover, we explicitly took into account the possible deformability of the inclusions and their surroundings by macroscopic variables. For this purpose, different strain tensors were introduced to describe the possibly different deformational states of these material components. A reformulation of this kind of description revealed \textit{relative strains} as a new macroscopic variable. 

During our motivation, we referred to polymeric materials as an illustration. There, situations are conceivable in which the dynamics of the inclusions or their immediate environment takes place on similar time scales as the macroscopic dynamics. Moreover, interpenetrating polymer networks may likewise be addressed by our approach. Yet, our derivation was based on symmetry principles only, not on specific chemical interaction details. Therefore, it should apply to any other composite material featuring the necessary match in time scales. Finally, if this match is not found and a separation in time scales prevails, 
our approach does not necessarily lose its significance completely. In that case, when the systems deform reversibly and are exposed to (quasi)static fields, the static part of our theory based on the presented variables combined to an energy density may still be meaningful. It reflects the additionally important inner degrees of freedom in composite materials.

\begin{acknowledgments}
The author thanks the Deutsche Forschungsgemeinschaft for support through the priority program SPP 1681 (no.\ ME~3571/3-2). 
\end{acknowledgments}
\vspace{-.0cm}


\end{document}